\documentclass{article}

\usepackage{graphicx}
\usepackage{subfigure}
\usepackage{booktabs} % for professional tables

\usepackage{hyperref}

\usepackage[accepted]{icml2021}

\usepackage{url}            % simple URL typesetting\
\usepackage{amsfonts}       % blackboard math symbols
\usepackage{nicefrac}       % compact symbols for 1/2, etc.
\usepackage{amsmath,amsfonts,amssymb}
\usepackage{algorithmic}
\usepackage{xcolor}
\usepackage{multirow}
\usepackage{enumerate}
\usepackage{tabularx}
\usepackage{comment}
\usepackage{dsfont}
\usepackage{multirow}
\usepackage[nameinlink,capitalize]{cleveref}
\usepackage{wrapfig}

\newcommand{\fw}[1]{{\color{purple}\bf [FW: #1]}}
\newcommand{\cg}[1]{{\color{brown}\bf [CG: #1]}}
\newcommand{\rk}[1]{{\color{green}\bf [RK: #1]}}

\newcommand{\tabincell}[2]{\begin{tabular}{@{}#1@{}}#2\end{tabular}}

\newcommand{\bw}{\mathbf{w}}

\newcommand{\by}{\mathbf{y}}
\newcommand{\bz}{\mathbf{z}}

\newcommand{\calM}{\mathcal{M}}
\newcommand{\calP}{\mathcal{P}}
\newcommand{\calY}{\mathcal{Y}}

\icmltitlerunning{Making Paper Reviewing Robust to Bid Manipulation Attacks}

\begin{document}

\twocolumn[
\icmltitle{Making Paper Reviewing Robust to Bid Manipulation Attacks}



\icmlsetsymbol{equal}{*}
\icmlsetsymbol{atcor}{\textdagger}

\begin{icmlauthorlist}
\icmlauthor{Ruihan Wu}{equal,cor}
\icmlauthor{Chuan Guo}{equal,fb}
\icmlauthor{Felix Wu}{asapp,atcor}
\icmlauthor{Rahul Kidambi}{amzn,atcor}
\icmlauthor{Laurens van der Maaten}{fb}
\icmlauthor{Kilian Q. Weinberger}{cor}
\end{icmlauthorlist}

\icmlaffiliation{cor}{Department of Computer Science, Cornell University}
\icmlaffiliation{fb}{Facebook AI Research}
\icmlaffiliation{asapp}{ASAPP}
\icmlaffiliation{amzn}{Amazon Search \& AI}

\icmlcorrespondingauthor{Ruihan Wu}{rw565@cornell.edu}
\icmlcorrespondingauthor{Chuan Guo}{chuanguo@fb.com}

\icmlkeywords{Machine Learning, ICML}

\vskip 0.3in
]



\printAffiliationsAndNotice{\icmlEqualContribution} % otherwise use the standard text.

\begin{abstract}
%!TEX root=../aistats_main.tex
Most computer science conferences rely on paper bidding to assign reviewers to papers. 
%Paper bidding allows these conferences to maintain high-quality assignments as they scale to (tens of) thousands of papers and reviewers.
Although paper bidding enables high-quality assignments in days of unprecedented submission numbers, it also opens the door for dishonest reviewers to adversarially influence paper reviewing assignments.
Anecdotal evidence suggests that some reviewers bid on papers by ``friends'' or colluding authors, even though these papers are outside their area of expertise, and recommend them for acceptance without considering the merit of the work.
In this paper, we study the efficacy of such \emph{bid manipulation attacks} and find that, indeed, they can jeopardize the integrity of the review process.
We develop a novel approach for paper bidding and assignment that is much more robust against such attacks.  
We show empirically that our approach provides robustness even when dishonest reviewers collude, have full knowledge of the assignment system's internal workings, and have access to the system's inputs. 
In addition to being more robust, the quality of our paper review assignments is comparable to that of current, non-robust assignment approaches.

\end{abstract}

%!TEX root=../icml_main.tex
\section{Introduction}

Peer review is a cornerstone of scientific publishing.
It also functions as a gatekeeper for publication in top-tier computer-science conferences.
%The underlying process crucially requires objectivity, unbiasedness, and transparency.
To facilitate high-quality peer reviews, it is imperative that paper submissions are reviewed by qualified reviewers.
In addition to assessing a reviewer's qualifications based on their prior publications~\citep{charlin2013toronto}, many conferences implement a \emph{paper bidding} phase in which reviewers express their interest in reviewing particular papers.
Facilitating bids is important because the review quality is higher when reviewers are interested in a paper~\citep{stent2018naacl}.

Unfortunately, paper bidding also creates the potential for difficult-to-detect adversarial behavior by reviewers.
In particular, a reviewer may place high bids on papers by ``friends'' or colluding authors, even when those papers are outside of the reviewer's area of expertise, with the purpose of accepting the papers without merit.
Anecdotal evidence suggests that such \emph{bid manipulation attacks} may have, indeed, influenced paper acceptance decisions in recent top-tier computer science conferences~\citep{vijaykumar2020potential}.

This paper investigates the efficacy of bid manipulation attacks in a realistic paper-assignment system.
We find that such systems are, indeed, very vulnerable to adversarial bid, which is corroborated by prior work~\cite{jecmen2020mitigating}.
Furthermore, we design a paper-assignment system that is robust against bid manipulation attacks.
Specifically, our system treats paper bids as supervision for a model of reviewer preferences, rather than directly using bids to assign papers.
We then detect atypical patterns in the paper bids by measuring their influence on the model, and remove such high-influence bids as they are potentially malicious.

We evaluate the efficacy of our system on a novel, synthetic dataset of paper bids and assignments that we developed to facilitate the study of robustness of paper-assignment systems.
We carefully designed this dataset to match the statistics of real bidding data from recent computer-science conferences.
We find that our system produces high-quality paper assignments on the synthetic dataset, while also providing robustness against groups of colluding, adversarial reviewers in a \emph{white-box setting} in which the adversaries have full knowledge of the system's inner workings and its inputs.
We hope our findings will help computer-science conferences in performing high-quality paper assignments at scale, while also minimizing the surface for adversarial behavior by a few bad actors in their community.

\begin{figure*}[ht!]
\centering
\includegraphics[width=0.9\linewidth]{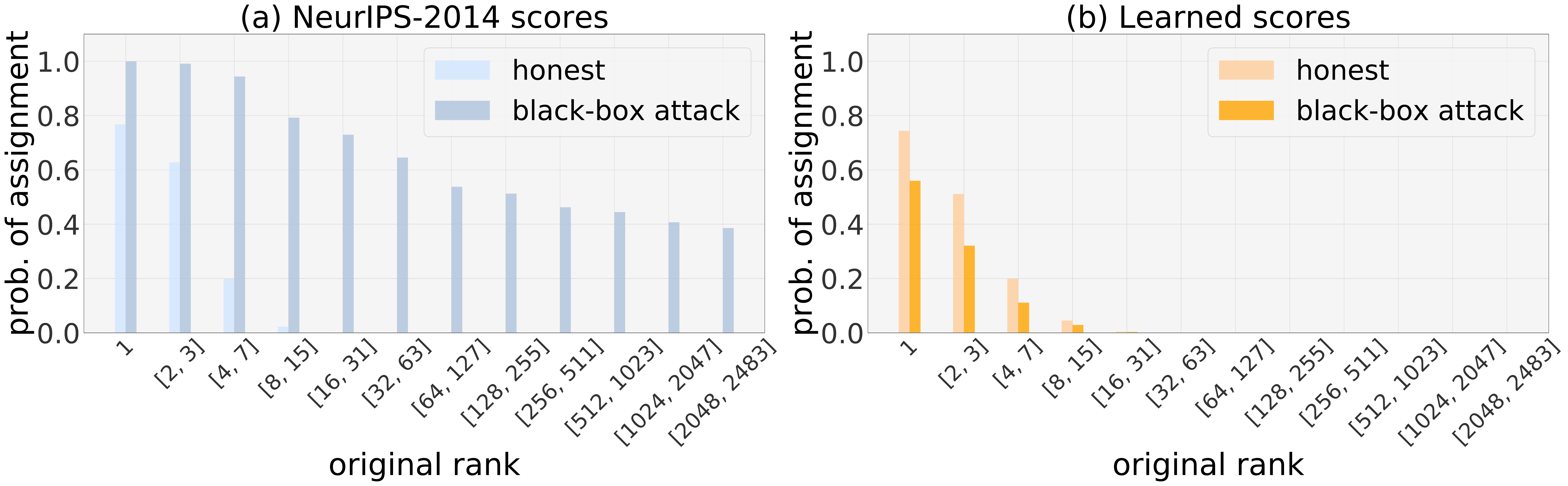}
\vspace{-3ex}
\caption{Probability of assigning an adversarial reviewer to the target paper before and after the reviewer executes their black-box bid manipulation attack. See text for details.}
\label{fig:prob_motivate}
\vspace{-1ex}
\end{figure*}

\section{Bid Manipulation Attacks}
\label{sec:motivation}
We start by investigating the effectiveness of bid manipulation attacks on a typical paper assignment system.

%We motivate our study into this security threat by demonstrating that it is feasible to attack a typical assignment system. Anecdotal evidence suggests that such attacks have been observed in practice. %We first describe the assignment system we consider.

{\bf Paper assignment system.}
Most paper assignment systems utilize a computed score $s_{r,p}$ for each reviewer-paper pair $(r,p)$ that reflects the degree of relevance between the reviewer and the paper~\citep{hartvigsen1999conference, goldsmith2007ai, tang2012optimization, charlin2013toronto}. The conference organizer can then maximize utility metrics such as the total relevance score whilst maintaining appropriate balance constraints: \emph{i.e.}, there are an adequate number of, say, $R$ reviewers per paper and every reviewer receives a manageable load of at most $P$ papers. This approach gives rise to the following optimization problem:
%\begin{equation}
%\sum_{r=1}^m \sum_{p=1}^n a_{r,p} s_{r,p},
%\label{eq:relevance}
%\end{equation}
%where $a_{r,p} \in \{0,1\}$ is the indicator variable of assigning a reviewer $r \in \{1,\ldots,m\}$ to a paper $p \in \{1,\ldots,n\}$. To ensure that each paper receives an adequate number of $R$ reviewers and each reviewer is given a sensible load of at most $P$ papers, we introduce additional balance constraints to arrive at the following optimization problem:
\begin{align}
\label{eq:assignment_prob}
\max_{a \in \{0,1\}^{m \times n}} \quad &\sum_{r=1}^m \sum_{p=1}^n a_{r,p} s_{r,p} \\
\text{subject to} \quad &\sum_{r=1}^m a_{r,p} = R \quad \forall p, \quad \sum_{p=1}^n a_{r,p} \leq P \quad \forall r, \nonumber
\end{align}
where $m$ and $n$ refer to the total number of reviewers and papers, respectively. ~\cref{eq:assignment_prob} is an assignment problem that can be solved using standard techniques such as the Hungarian algorithm~\citep{kuhn1955hungarian}.

The reviewer-paper relevance score, $s_{r,p}$, is critical in obtaining high-quality assignments. Arguably, an ideal relevance score incorporates both the reviewer's \emph{expertise} and \emph{interest} towards the paper~\citep{stent2018naacl}. Approaches for measuring expertise include computing the similarity of textural features between reviewers and papers~\citep{dumais1992automating, mimno2007expertise, charlin2013toronto} as well as using authorship graphs~\citep{rodriguez2008algorithm, liu2014robust}. In addition to these features, paper assignment systems generally consider reviewer interest obtained via self-reported paper bids. For example, the NeurIPS-2014 assignment system \citep{lawrence2014nips} uses a formula for $s_{r,p}$ that incorporates the reviewer's and paper's subject area, TPMS score~\citep{charlin2013toronto}, and the reviewer's bid. Each reviewer may bid on a paper as \textit{none}, \textit{in a pinch}, \textit{willing}, or \textit{eager}\footnote{For simplicity, we exclude the option \textit{not willing} that expresses negative interest.} to express their preference. The {\em none} option is the default bid when a reviewer did not enter a bid.

{\bf Bid manipulation attacks.}
Although incorporating reviewer interest via self-reported bids is beneficial to the overall assignment quality, it also allows a malicious reviewer to bid \emph{eager} on a paper that is outside their area of expertise, with the sole purpose of influencing the acceptance decision of a paper that was authored by a ``friend'' or a ``rival''.
If a single bid has too much influence on the overall assignment, such bid manipulation attacks may be effective and jeopardize the integrity of the review process.

We demonstrate the feasibility of a simple \emph{black-box} bid manipulation attack against the assignment system in~\cref{eq:assignment_prob}. For a target paper $p$, the malicious reviewer attacks the assignment system by bidding \emph{eager} for $p$ and \emph{none} for all other papers.
%The attack strategy is as follows: For a target paper $p$, the malicious reviewer bids \textit{eager} for $p$ and \textit{none} (\emph{i.e.}, no bid) for all other papers.
We evaluate the effectiveness of the attack by randomly picking 400 papers from our synthetic conference dataset (see \cref{sec:experiment}), and determine paper assignments using~\cref{eq:assignment_prob} (with $R=3$ and $P=6$) using relevance scores from the NeurIPS-2014 system~\citep{lawrence2014nips}.
\cref{fig:prob_motivate} (\emph{left}) shows the fraction of adversarial reviewers ($m=2,483$) that can secure their target paper in the final assignment via the bid manipulation attack. As an attack is easier if a reviewer is already ranked high for a particular paper (\emph{e.g.}, because nobody else bids on this paper, or the subject areas match), we visualize the success rate as a function of rank of the ``true'' paper-reviewer relevance score. More precisely, we rank all reviewers by their original (pre-manipulation) relevance score $s_{r,p}$ and group them into bins of increasing size.
%
% For each paper we rank the reviewers by their genuine relevance scores $s_{r,p}$ and test if a reviewer can succeed at getting the paper assigned
%  For visualization purposes, we bin the reviewers by rank.
%$[1,m]$ into exponentially-spaced bins $[1], [2,3], [4,7],\ldots$ We execute the attack for each reviewer and produce assignments .
%

The light gray bar in each bin reports the assignment success rate if all reviewers bid honestly. In the absence of malicious reviewers, the majority of assignments go to reviewers ranked 1 to 7. However, with malicious bids, {\em any} reviewer stands a good chance of being assigned the target paper. For instance, the chance of getting a target paper for a reviewer ranked between 16 and 31 increases from 0\% to over 70\% when bidding maliciously. Even reviewers with the lowest ranks (2048 and lower) have a 40\% chance of being assigned the target paper by just changing their bids. This possibility is especially concerning because it may be much easier for an author to corrupt a non-expert reviewer (\emph{i.e.}, a reviewer with a relatively low rank), simply because there are many more such reviewer candidates. %This attack shows that the scoring and assignment function is strongly susceptible to changes in the reviewers' bids.

%This security loophole exposes an inherent tension within the assignment process. On the one hand, assigning papers that a reviewer has indicated specific interest towards naturally helps in eliciting meaningful feedback; whilst an over-reliance on the reviewers preferences leads to potential for such assignment systems to be susceptible for allowing academically dishonest behavior.

%The root of this security loophole that allows the blatant violation of academic integrity lies in the incorporation of self-reported interest. However, it arguable that accounting for reviewer interest is beneficial for receiving a high quality of reviews, and thus, this paper seeks to utilize the bidding information in a more robust manner.%we would like to leverage this information in a more robust manner.

%!TEX root=../icml_main.tex
\section{Predicting Relevance Scores}
\label{sec:model}

The success of the bid manipulation attack exposes an inherent tension in the assignment process. Assigning papers to a reviewer who has expressed explicit interest helps in eliciting high-quality feedback. However, relying too heavily on individual bids paves the way for misuse by malicious reviewers. To achieve a better trade-off, we propose to use the bids from all reviewers (of which the vast majority are honest) as labels to train a supervised model that \emph{predicts} bids as the similarity score $s_{r,p}$, and all other indicators (\emph{e.g.}, subject area matches, TPMS score~\citep{charlin2013toronto}, and paper title) as features.
This indirect use of bids allows the scoring function to capture reviewer preferences but reduces the potential for abuse. Later, we will show that this approach also allows for the development of active defenses against bid manipulation attacks.

%fills the bidding matrix
%discover unreported interest and improve the assignment quality for papers that had few bids. For malicious reviewers whose bidding behavior is abnormal, this score can correct for the corrupted bids by learning the general bidding pattern from the majority of honest reviewers.

%The purpose of our scoring model is to learn a mapping from reviewer-paper pairs to relevance scores that generalizes to unobserved reviewer preferences. We assume the vast majority of reviewers are well intentioned, and there may exist a small subset of reviewers who maybe malicious and exhibit anomalous bidding patterns. Thus, as we shall see below, the supervised learning view admits the potential of learning the typical bidding behavior of an honest reviewer.

\noindent\textbf{Scoring model.} Let $X \in \mathbb{R}^{(mn) \times d}$ be a feature matrix consisting of $d$-dimensional feature vectors for every pair of $m$ reviewers and $n$ papers. Let $\calY$ denote the set of possible bids in numerical form, \emph{e.g.} $\calY=\{0,1,2,3\}$. We define $\by\in\calY^{mn}$ as the label vector containing the numerical bids for all reviewer-paper pairs. We define a ridge regressor that maps reviewer-paper features to corresponding bids, similar to the linear regression model from~\citet{charlin2013toronto}:
%We assign real-valued scores from a label set $\calY \subseteq \mathbb{R}_{\geq 0}$ to the bids and regress the feature matrix $X$ against the training label vector $\by \calY^{mn}$ that with $\mathbf{y}_{rp}$
\begin{equation}
    \mathbf{w}^* = \mathrm{argmin}_{\mathbf{w}}\| X\mathbf{w} - \mathbf{y} \|_2^2 + \lambda \|\mathbf{w}\|_2^2.
    \label{eq:ridge_regression}
\end{equation}

To ensure that no single reviewer has disproportionate influence on the model, we restrict the maximum number of positive bids from a reviewer to be at most $U=60$ and subsample bids of a reviewer whenever the number of bids exceeds $U$. In a typical CS conference, most reviewers bid on no more than 60 papers (out of thousands of submissions)~\cite{shah2018design}.

The trained model $\bw^*$ can predict reviewer interest by computing a score $s_{r,p}$ for a reviewer-paper pair $(r,p)$ as follows:
\begin{equation}
s_{r,p} = X_{r,p} \bw^* = X_{r,p} H^{-1} X^\top \by,
\label{eq:relevance}
\end{equation}
where $H = X^\top X + \lambda I$ is the ridge Hessian (size $d\times d$) and $X_{r,p}$ is the feature vector for the pair $(r,p)$. These predicted scores can then be used in the assignment algorithm in place of bids. In Appendix \ref{sec:features}, we validate the prediction accuracy of our model using the average precision-at-k (AP@k) metric.

There is an important advantage to our method: bidding is a laborious and monotonous task, and as mentioned above most reviewers only bid on very limited papers. It is likely that only a partial set of bids is observed among all papers that the reviewer is interested in. The scoring model could fill in missing scores by learning the latent interest from the features of papers and reviewers. Completing the full bidding matrix improves the assignment quality, particularly for papers that received few bids originally.

The choice of regression loss serves an important purpose. Since the bid value (between 0 and 3) reflects the \emph{degree of interest} from a reviewer, the loss should reflect the severity of error when making a wrong prediction.
%We'd like to highlight the reasonability of regression setting rather than classification setting. Treating the bids as categorical labels is unsuitable because the value (between 0 and 3) reflects the \emph{degree} of interest.
For example, if a reviewer expresses \emph{eager} interest (bid score 3), predicting \emph{no bid} (bid score 0) would incur a much greater loss than predicting \emph{willing} (bid score 2).

\noindent\textbf{Effect against simple black-box attack.} \cref{fig:prob_motivate} (\emph{right}) shows the effect of the proposed scoring model against the bid manipulation attack from \cref{sec:motivation}. The assignment probability for honest bidders (light orange) is similar to that of the NeurIPS-2014 system across different bins of reviewer rank. However, deviations from benign bidding behavior are clearly corrected by the model: in fact, the assignment probability \emph{decreases} after the attack (dark orange).
This can be explained by the fact that our approach does not use bids to assign  reviewers to papers directly, but instead to learn for what type of papers a reviewer may be suitable.
The reviewer is actually well-suited for high ranking submissions, but by only bidding on the target paper (instead of honest bids on similar submissions) the model receives less signal that suggests the reviewer is a match for the target paper.
%and their honest bidding would substantiate this fact. .

%!TEX root=../icml_main.tex
\section{Defending Against Colluding Bid Manipulation Attackers}
\label{sec:colluding}

Although the learning-based approach appears robust against manipulation of bids by one reviewer, attackers may have stronger capabilities. Specifically, an adversary can modify their bids based on knowledge of a friend/rival's submissions or another reviewer's bids. Moreover, adversarial reviewers may \emph{collude} to secure the assignment of a specific paper.
We capture such capabilities in a \emph{threat model} that describes our assumptions about the adversary.
We design an optimal \emph{white-box attack} in this threat model that drastically improves the adversary's success rate.
Both the threat model and the white-box attack are intentionally designed to provide very broad capabilities to the adversary.
Next, we design a defense that detects and removes white-box adversaries from the reviewer pool to provide security even under the new threat model.
%We demonstrate that the bidding mechanism introduces \emph{no additional} security risk under our defense.

%To address these issues, we first detail our assumptions about the capabilities of an adversary. Under these assumptions, we design an optimal \emph{white-box} attack that drastically improves the adversary's success rate against the scoring-based assignment framework. Despite the unrealistic capabilities for the colluding adversary, we design a powerful defense strategy that detects and removes potential attackers from the reviewer pool and demonstrate its efficacy against the proposed white-box attack.

\paragraph{Threat Model.}
We make the following assumptions about adversarial reviewers:
%\begin{enumerate}[1.]
\textbf{1.} The adversary may collude with one or more reviewers to secure a target paper's assignment.\footnote{\emph{e.g.} by posting the paper ID in a private chat channel of college alumni or like minded members of the community.} If any of the colluding reviewers are assigned the paper in question, the attack is considered successful. Collusion with \emph{any} reviewer is allowed except the top-ranked candidates (based on honest bidding), as this would not be an abuse of the \emph{bidding} process\footnote{For this reason, our framework is not suitable for preventing the attack in~\citep{vijaykumar2020potential} since collusion likely occurred in the author stage.}.
%This situation represents an organized effort of authors and/or reviewers. %The reviewer may collude with one or more other reviewers and jointly attempt to secure the target paper's assignment. An attack is deemed successful if either she or any reviewer in her colluding party are assigned the paper in question. The collusion can arise from a mutual agreement between several authors or from an organizational effort by a perpetrator.
%
\textbf{2.} The adversary cannot manipulate any training features. We are interested in preventing against the \emph{additional} security risk enabled by the bidding mechanism. An attack that succeeds by manipulating features can also be used against an automated assignment system that does not allow bidding.
\textbf{3.} The adversary may have full knowledge of the assignment system.
\textbf{4.} The adversary may have direct access to the features and bids of all other reviewers. %That is, we assume the most pessimistic scenario of a \emph{white-box attack}, where the reviewer and his/her colluders can access all reviewers' features and bids.
\textbf{5.} The adversary may be able to arbitrarily manipulate his/her bids and those of anyone in the colluding group.

\subsection{White-box Attack}
\label{sec:cheating}
To successfully attack the assignment system under these assumptions, the adversary needs to maximize the predicted relevance score of the target paper for him/herself and/or the other colluding reviewers.
This amounts to executing a data poisoning attack~\citep{biggio2012poisoning, xiao2015support, mei2015using, jagielski2018manipulating, koh2018stronger} against the regression model that is used to predict scores, aiming to alter the score prediction for a specific paper-reviewer pair.

{\bf Non-colluding attack.} We first devise an attack that maximizes the malicious reviewer's score $s_{r,p}$ for target paper $p$ in the non-colluding setting. We represent reviewers as $[m] = \{1,\ldots,m\}$ and let
$$\calY_\text{feas} = \{\by' \in \calY^n : |\{q: \by'_q > 0\}| \leq U\}$$
denote the feasible set of bidding vectors for a particular reviewer for which the number of positive bids is at most $U$. Adversary $r$ can change $\by_r$ to the $\by_r' \in \calY_\text{feas}$ that maximizes the relevance score:
\vspace{-2ex}
\begin{align*}
s_{r,p}^* &:= \max_{\by_r' \in \calY_\text{feas}} X_{r,p} H^{-1} (X_r^\top \by_r' + X_{[m] \setminus \{r\}}^\top \by_{[m] \setminus \{r\}}) \\
&= \max_{\by_r' \in \calY_\text{feas}} X_{r,p} H^{-1} (X_r^\top \by_r' - X_r^\top \by_r + X^\top \by).
\end{align*}
It is straightforward to see that $s_{r,p}^*$ maximally increases the score prediction for reviewer $r$:
\begin{equation}
\Delta s_{r,p}^* := s_{r,p}^* - s_{r,p} = \max_{\by_r' \in \calY_\text{feas}} X_{r,p} H^{-1} X_r^\top (\by_r' - \by_r).
\label{eq:wb_obj_single}
\end{equation}
Note that \cref{eq:wb_obj_single} maximizes the inner product between $\bz := X_{r,p} H^{-1} X_r^\top$ and $\by_r' - \by_r$. To achieve the maximum, papers $q$ corresponding to the top-$U$ positive values in $\bz$ should be assigned $\by_{r,q} = \max \calY$, and the remaining bids are set to 0. This requires the adversary to solve a top-$U$ selection problems, which can be done in $O(d^2 + n (d + \log U))$~\citep{clrs2009}.

{\bf Colluding attack.} Adversarial reviewers can collude to more effectively maximize the predicted score for reviewer $r$. An attack in this setting maximizes over the colluding group, $\calM$, and over the bids of every reviewer in $\calM$. We note that \cref{eq:wb_obj_single} is not specific to reviewer $r$, but that the influence of any reviewer $t$'s bids on score prediction $s_{r,p}$ has the form: $$\Delta_t s_{r,p} := \max_{\by_t' \in \calY_\text{feas}} X_{r,p} H^{-1} X_t^\top (\by_t' - \by_t).$$ Hence, the influence from the members of $\calM$ on $s_{r,p}$ are {\em independent}, which implies the adversaries can adopt a greedy approach.
Specifically, $M_a$ colluding adversaries can alter the $(M_a n)$-dimensional training label vector $\by_{\calM}$ to $\by_{\calM}' \in \calY_\text{feas}^{M_a}$ to maximize the score prediction for reviewer $r$ via:
\begin{align}
\Delta s_{r,p}^* &= \max_{(\calM, \by_{\calM}') \in \calP(r, M_a)} X_{r,p} H^{-1} X_\calM^\top (\by_\calM' - \by_\calM),\nonumber \\
&= \max_{\calM \subseteq [m] : r \in \calM, |\calM|=M_a} \sum_{t \in \calM} \max_{\by_t' \in \calY_\text{feas}} X_{r,p} H^{-1} X_t^\top (\by_t' - \by_t) \nonumber \\
&= \max_{\calM \subseteq [m] : r \in \calM, |\calM|=M_a} \sum_{t \in \calM} \Delta_t s_{r,p},
\label{eq:wb_greedy}
\end{align}
where $\calP(r, M_a)$ denotes the set of possible colluding parties of size $M_a$ and their bids:
\begin{align*}
\calP(r, M_a) := \{ (\calM, \by_\calM') : \calM \subseteq [m], &r \in \calM, |\calM| = M_a \\
&\text{ and } \by_{\calM}' \in \calY_\text{feas}^{M_a}\}.
\end{align*}
The last line in \cref{eq:wb_greedy} can be computed by first evaluating $\Delta_t s_{r,p}$ for every $t \in [m] \setminus \{r\}$, and then greedily selecting the top-$(M_a-1)$ reviewers to form the colluding party with $r$. The computational complexity of the resulting attack is $O(d^2 + mn (d + \log U) + m \log M_a))$.

%Note that the second line follows since the dropped term doesn't affect the maximization. Observe that for each $t \in \calM$, the term $X_{r,p} H^{-1} X_t^\top (\by_t' - \by_t)$ is precisely the influence function~\citep{cook1982residuals, koh2017understanding} of reviewer $t$ on $s_{r,p}$. In other words, this white-box attack seeks to find the colluding party that can maximally influence the predicted score for $(r,p)$. The second term in Equation \ref{eq:wb_greedy} can be computed greedily and the computational complexity is $O(d^2 + mn (d + \log U) + m \log M_a))$.

%\paragraph{Relationship to influence function.} The white-box attack defined above is closely related to a well-studied statistical quantity known as the influence function~\citep{cook1982residuals, koh2017understanding}, which measures the change in prediction resulting from a change in the input. For the linear/ridge regression relevance scoring model, the influence on the score for $(r,p)$ from a change of bids to $\by_{r'}$ by a reviewer $r'$ is: $$(\Delta s_{r,p})_{r'} = X_{r,p} H^{-1} X_{r'}^\top (\by_{r'} - \by_{r'}),$$ which is precisely the contribution of a reviewer $r' \in \calM$ in Equation \ref{eq:wb_greedy}. Thus, the white-box attack we defined is equivalent to finding the colluding party $\calM$ that can maximally influence the predicted score for $(r,p)$.

\subsection{Active Defense Against Colluding Bid Manipulation Attacks}
\label{sec:detection}

Both the black-box attack from \cref{sec:motivation} and the white-box attack described above adversarially manipulate paper bids. In contrast to honest reviewers whose bids are strongly correlated with their expertise and subject of interest, attackers provide ``surprising'' bids that have a large influence on the predictions of the scoring model.
This allows us to detect potentially malicious bids using an outlier detection algorithm.
Specifically, we make our paper assignment system robust against the colluding bid manipulation attacks by detecting and removing training examples that have a disproportional influence on model predictions.
We make the same assumptions about the attacker as in \cref{sec:cheating}, and, in addition, that they are unaware of our active defense.

%We develop a paper assignment system that is robust against the colluding label-poisoning attack described above by detecting and removing training points that have a disproportional influence on model predictions.
%The system first identifies a \emph{candidate set} that contains only the top-$K$ reviewers for each paper.
%This hardly affects the assignments since low-ranking reviewers for a paper are rarely assigned that paper.
%Next, we identify the subset of reviewers that would \emph{reduce the predicted relevance scores the most they were not present} by searching over the candidate set of reviewers.
%Finally, we re-compute the relevance scores without considering the potentially malicious reviewers, and perform the assignment using the re-computed scores.

To implement this system, we note that given a set of malicious reviewers $\calM$, we can re-compute the relevance scores for a reviewer-paper pair $(r,p)$ by removing these reviewers from the training set: $$\tilde{s}_{r,p} = X_{r,p} H_{\calM^c}^{-1} X_{\calM^c}^\top \by_{\calM^c},$$ where $H_{\calM^c} = X_{\calM^c}^\top X_{\calM^c} + \lambda I$ is the Hessian matrix for data points in the complement of the malicious reviewer set $\calM$.
We assume that at most $M_d$ reviewers collude to form set $\calM$.
Intuitively, $\tilde{s}_{r,p}$ reflects the relevance score for the pair $(r,p)$ \emph{as predicted by other reviewers}. Relying on the assumption that the vast majority of reviewers are benign, $\tilde{s}_{r,p}$ is likely close to the unobserved true preferences had $r$ been benign.

\begin{algorithm}[!t]
\begin{algorithmic}[1]
 \STATE Predict relevance scores $s_{r,p}$ for all reviewer-paper pairs;
 \STATE Initialize candidate set $C = \{(r,p) : \text{rank}(s_{r,p}) \text{ is at least } K \text{ for paper } p\}$;
 \FOR{reviewer-paper pair $(r,p) \in C$}
 \STATE Compute relevance score $s_{r,p}^\dagger$ using \cref{eq:detect_approx} \;
 \STATE Remove $(r,p)$ from $C$ if $\text{rank}(s_{r,p}^\dagger)$ is below $K$ for paper $p$;
 \ENDFOR
 \STATE Solve the assignment problem in \cref{eq:assignment_prob} using $s_{r,p}$ for pairs in $C$.
\end{algorithmic}
\caption{Paper assignment system that is robust against colluding bid manipulation attacks.}
\label{alg:framework}
\end{algorithm}

Following work on robust regression~\citep{jagielski2018manipulating,chen2013robust,bhatia2015robust}, this allows us to compute relevance scores that ignore the most likely malicious reviewers in $\calM$ by evaluating:
\begin{equation}
    s_{r,p}^\dagger = \min_{\calM \subseteq [m]: r \in \calM, |\calM|=M_d} X_{r,p} H_{\calM^c}^{-1} X_{\calM^c}^\top \by_{\calM^c} \leq \tilde{s}_{r,p}.
    \label{eq:removal_opt}
\end{equation}
That is, $s_{r,p}^\dagger$ overestimates the decrease in the predicted relevance score for $(r,p)$ had $r$ been benign. The optimization problem in \cref{eq:removal_opt} is intractable because it searches over ${m-1 \choose M_d-1} = \Theta(m^{M_d})$ subsets of reviewers, $\calM$, and because it inverts a $d \times d$ Hessian for every $\calM$. %The sheer size of the enumeration set coupled with matrix inversion computations makes this problem infeasible. %Even when using the Woodbury inversion formula to perform a low-rank update of the full Hessian $H^{-1}$, the computational complexity is still $\Theta(M_d n d^2)$ per inversion. Thus the total cost for computing $s_{r,p}^\dagger$ for each of the $M_d n$ pairs in the candidate set is $\Theta(m^{M_d} M_d n d^2)$, which is certainly infeasible.
To make optimization tractable, we approximate the Hessian $H_{\calM^c}^{-1}$ by $\color{blue}{H^{-1}}$, which is accurate for small $M_d$. This approximation facilitates a greedy search for $\calM$ because it allows \cref{eq:removal_opt} to be decomposed:
\begin{align}
s_{r,p}^\dagger &\approx \min_{\calM \subseteq [m]: r \in \calM, |\calM|=M_d} X_{r,p} {\color{blue} H^{-1}} X_{\calM^c}^\top \by_{\calM^c} \nonumber \\
&= X_{r,p} H^{-1} X^\top \by - \nonumber \\
&\hspace{6ex} \max_{\calM \subseteq [m]: t \in \calM, |\calM|=M_d} \sum_{t \in \calM} X_{r,p} H^{-1} X_t \by_t.
\label{eq:detect_approx}
\end{align}
\cref{eq:detect_approx} can be computed efficiently by sorting the values of $S = \{ X_{r,p} H^{-1} X_t \by_t : t \neq r\}$ and selecting $r$ as well as the top $M_d - 1$ corresponding reviewers in $S$. The computational complexity of the resulting algorithm is $O(d^2 + mnd + m \log M_d))$ for each pair $(r,p)$.

{\bf Assignment algorithm.} Efficient approximation for the robust relevance score $s_{r,p}^\dagger$ enables our robust assignment algorithm, which proceeds as follows.
We first form the \emph{candidate set} $C$ of reviewer-paper pairs by selecting the top-$K$ reviewers for each paper according to the predicted relevance score $s_{r,p}$.
For each pair $(r,p) \in C$, the algorithm marks $r$ as potentially malicious and removes the pair $(r,p)$ from $C$ if $r$ \emph{would not have belonged to the candidate set} using the robust relevance score $s_{r,p}^\dagger$.
Since $s_{r,p}^\dagger \leq \tilde{s}_{r,p}$, an $M_a$-colluding attack is always marked as malicious if $M_a \leq M_d$.
After removing every potentially malicious pair from $C$, the assignment problem in \cref{eq:assignment_prob} is solved over the remaining reviewer-paper pairs in the candidate set to produce the final assignment\footnote{This can be achieved by setting $s_{r,p} = -\infty$ for all $(r,p) \notin C$.}.
The resulting assignment algorithm is summarized in \cref{alg:framework}. The algorithm trades off two main goals:
%There are two desirable properties of the final assignment:
\vspace{-2ex}
\begin{enumerate}[1.]
\setlength\itemsep{0ex}
\item Every paper needs to be assigned to a sufficient number of reviewers that have the expertise and willingness to review. Therefore, the approach that removes potentially malicious reviewer candidates needs to have a low false positive rate (FPR).
\item The final assignment should be robust against collusion attacks. Therefore, the approach that filters out potentially malicious reviewers needs to have a high true positive rate (TPR).
\end{enumerate}
\vspace{-2ex}
This trade-off between FPR and TPR is governed by the hyperparameter $M_d$. Using a higher value of $M_d$ can provide robustness against larger collusions, but it may also remove many benign reviewers from the candidate set even when insufficient alternative reviewers are available. We perform a detailed study of this trade-off in \cref{sec:experiment}.

%!TEX root=../icml_main.tex
\section{Experiments}
\label{sec:experiment}

We empirically study the efficacy of our robust paper bidding and assignment algorithm.
Our experiments show that our assignment algorithm removes a large fraction of malicious reviewers, while still preserving the utility of bids for honest reviewers.

{\bf Dataset.}
Because real bidding data is not publicly available, we construct a synthetic conference dataset from the Semantic Scholar Open Research Corpus~\citep{ammar2018construction}.
This corpus contains publicly available academic papers annotated with attributes such as citation, venue, and field of study.
To simulate a NeurIPS-like conference environment, we collect $n=2446$ papers published in AI conferences between 2014 and 2015 to serve as submitted papers.
We also select $m=2483$ authors to serve as reviewers, and generate bids based on paper citations.
Generated bids are selected from the set $\calY = \{0,1,2,3\}$, corresponding to the bids \emph{none}, \emph{in a pinch}, \emph{willing}, and \emph{eager}.

We generated bids in such a way as to mimic bidding statistics from a recent, major AI conference.
Our paper and reviewer features include paper/reviewer subject area, paper title, and a TPMS-like similarity score. We refer to the appendix for more details on our synthetic dataset.
% suggested change from legal
%We did validate our findings on a real bidding dataset from a major computer science conference, which lead to identical conclusions. Unfortunately, we cannot include these results in this publication due to privacy concerns.
For full reproducibility we release our code\footnote{\url{https://github.com/facebookresearch/secure-paper-bidding}} and synthetic data\footnote{\url{https://drive.google.com/drive/folders/1khI9kaPy_8F0GtAzwR-48Jc3rsQmBhfe}} publicly and invite program chairs across disciplines to use our approach on their real bidding data.

\begin{comment}
\begin{equation}
\label{eq:wb_esb_obj}
    s_{r,p}^* := \max_{(\calM, \by_{\calM}') \in \calP(r, M_a)} \sum_{i=1}^3 X_{r,p}^{(i)}H^{-1}_{(i)} (X_\calM^{(i) \top} \by_\calM' + X_{\calM^c}^{(i) \top} \by_{\calM^c})
\end{equation}
and
\begin{equation}
    s_{r,p}^\dagger = \min_{\calM \subseteq \{1,\ldots,m\}: r \in \calM, |\calM|=M_d} \sum_{i=3}^3 X_{r,p}^{(i)} H_{(i)\calM^c}^{-1} X_{\calM^c}^{(i) \top} \by_{\calM^c}.
    \label{eq:removal_opt_esb},
\end{equation}
where $X^{(i)}$ and $H_{(i)}$ are the $i$th hashed features and corresponding hessian matrix. Both (\ref{eq:wb_esb_obj}) and (\ref{eq:removal_opt_esb}) can be solved in the same way as (\ref{eq:wb_obj}) and (\ref{eq:detect_approx}) respectively.

\paragraph{Assumptions for the dataset.}
We assume that the bidding scores in our dataset are all given by benign reviewers.
There is no cheating bid in our dataset initially.
This assumption helps us to describe the evaluation.
Even if this assumption is not true, results in this section still help understand both the cheating and detection processes.
\end{comment}

\subsection{Effectiveness of White-Box Attacks}
\label{sec:white-box}

\begin{comment}
\begin{wrapfigure}{r}{0.55\textwidth}
\centering
\vspace{-5ex}
\includegraphics[width=0.55\textwidth]{figs/prob_impr}
\vspace{-4ex}
\caption{Success rate after the white-box bid manipulation attack against an undefended linear regression scoring model. Initially, all reviewers are below rank $K=50$ and have no chance of being assigned. By manipulating their bids, colluding reviewers can drastically increase their success rate, especially with higher collusion size $M_a$.}
\vspace{-2ex}
\label{fig:prob}
\end{wrapfigure}
\end{comment}

\begin{figure}[t!]
\centering
\includegraphics[width=\columnwidth]{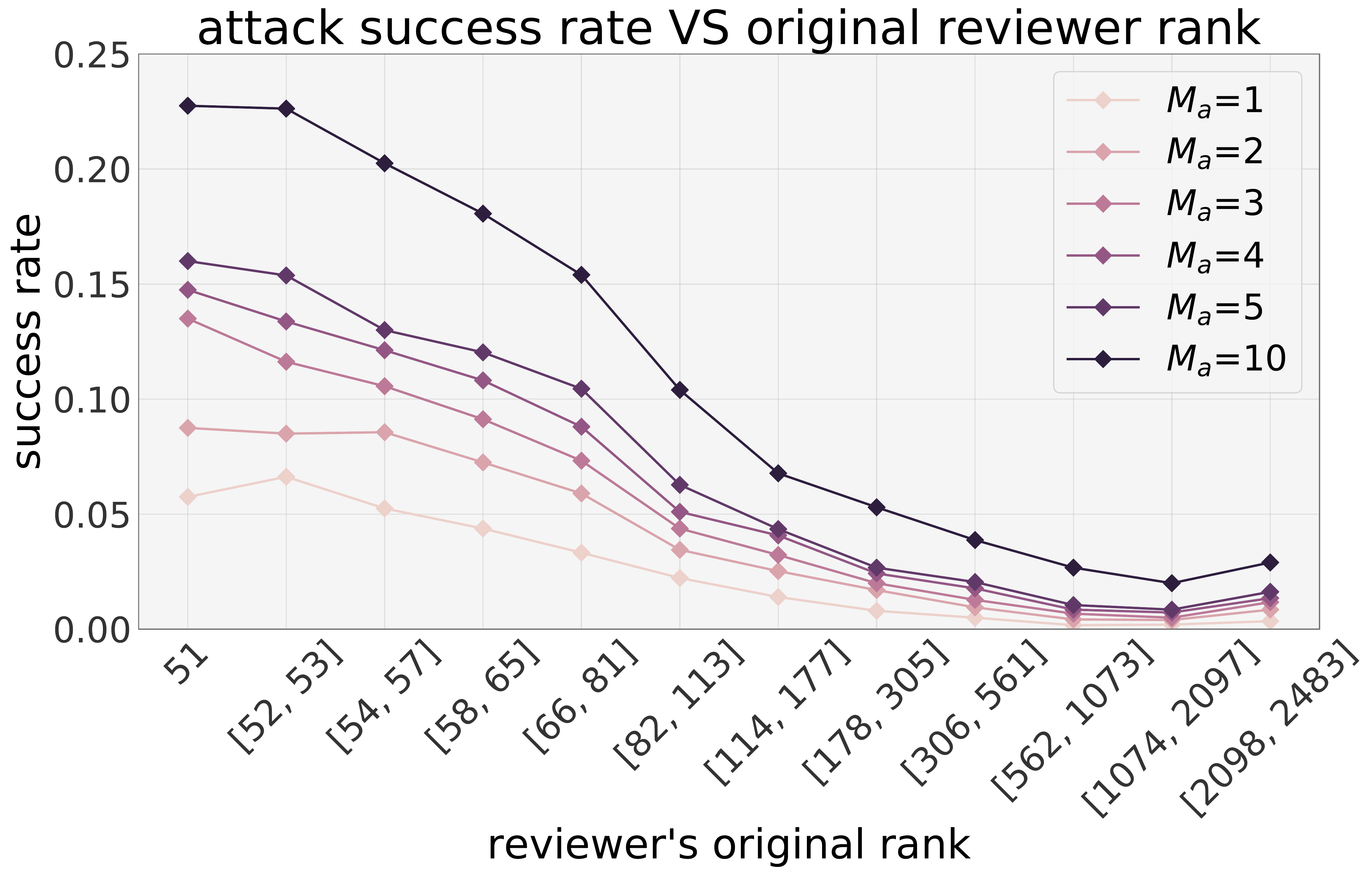}
\vspace{-5ex}
\caption{Success rate after the white-box bid manipulation attack against an undefended linear regression scoring model.} %Initially, all reviewers are below rank $K=50$ and have no chance of being assigned. By manipulating their bids, colluding reviewers can drastically increase their success rate, especially with higher collusion size $M_a$.}
\vspace{-3ex}
\label{fig:prob}
\end{figure}

We first show that the white-box attack from \cref{sec:cheating} can succeed against our relevance scoring model if detection of malicious reviewers is not used. We perform the white-box attacks as follows:\\
\textbf{1.} The relevance scoring model is trained to predict scores $s_{r,p}$ for every reviewer-paper pair.\\
\textbf{2.} We randomly select 400 papers and rank all $m=2483$ reviewers for these papers based on $s_{r,p}$.\\
\textbf{3.} We discard the $K=50$ highest-ranked reviewers as attacker candidates for paper $p$ because high-ranked reviewers need not act maliciously to be assigned.\\
\textbf{4.} We group the remaining reviewers into bins of exponentially growing size (powers of two), and sample 10 malicious reviewers from each bin without replacement.\\
\textbf{5.} Each selected reviewer chooses its most suitable $M_a$ colluders and modifies their bids using the attack from \cref{sec:cheating}, targeting paper $p$.

{\bf Result.}
We run our assignment algorithm on the maliciously modified bids and evaluate the chance of assignment for reviewer $r$ before and after the attack.
\cref{fig:prob} shows the fraction of malicious reviewers that successfully alter the paper assignments and is assigned their target paper. Each line shows the attack success rate with a certain colluding party size of $M_a$.
When bidding honestly, all reviewers are below rank $K=50$ and have no chance of being assigned.
With a colluding party size of $M_a=10$, a reviewer has a 22\% chance of being assigned the target paper at an original rank of 51. At the same rank, the success rate is up to 5\% even when no collusion occurs. Increasing the collusion size $M_a$ strictly increases the assignment probability, while attackers starting from a lower original rank have a lower success rate. The latter trend shows that the model provides a limited degree of robustness even without the detection mechanism.

\begin{figure}[t!]
\centering
\includegraphics[width=\columnwidth]{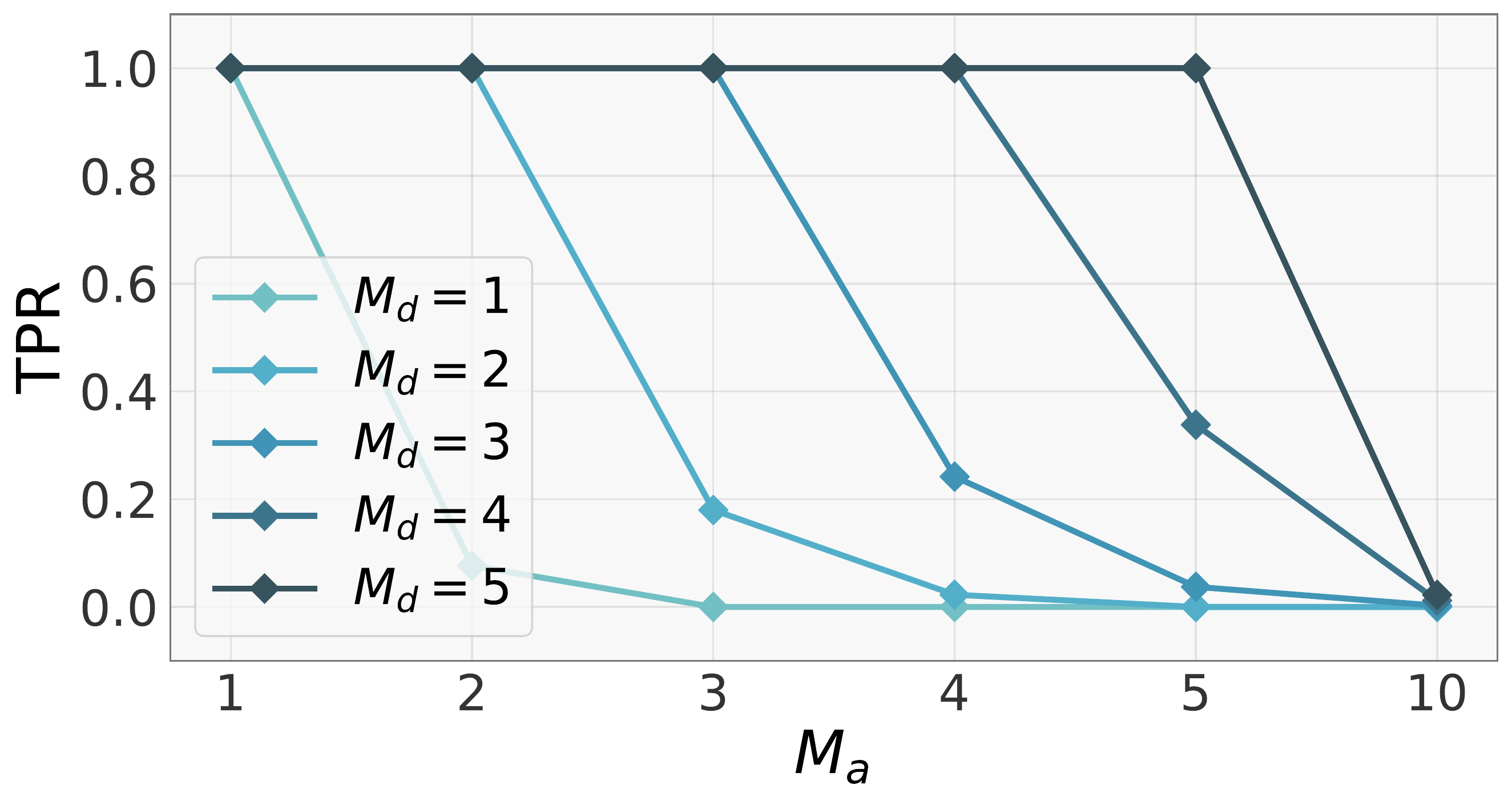}
\vspace{-4ex}
\caption{TPR for detecting successful \emph{white-box} attacks using \cref{alg:framework}. For colluding parties of size $M_a \leq M_d$, the detection algorithm has a \emph{near-perfect} TPR. Detection remains viable even when $M_a > M_d$ for moderately high values of $M_d$.}
\label{fig:tpr}
\vspace{-3ex}
\end{figure}

\begin{table*}[t!]
\small
	\centering
	\resizebox{\textwidth}{!}{%
	\begin{tabular}{c|cc|cccc|c}
		\toprule
		\bf Setting & \multicolumn{2}{c|}{\bf FPR} & \multicolumn{4}{c|}{\bf Assignment Quality} & \bf \# of under- \\
		%\cline{3-7}
	   & \bf Top-5 & \bf Top-50 & \bf Frac. of pos. & \bf Avg. bid score & \bf Avg. TPMS & \bf Avg. max. TPMS & \bf reviewed \\
		\midrule
    NeurIPS-2014 & -- & -- & 0.990 & 2.732 & 0.732 & 0.737 & -- \\
    TPMS only & -- & -- & 0.323 & 0.872 & 0.949 & 0.997 & -- \\
    \midrule
		$M_d=0$ & -- & -- & 0.442 & 1.200 & 0.848 & 0.943 & -- \\
		\midrule
		$M_d=1$ & 0.022 & 0.259 & 0.443 & 1.201 & 0.849 & 0.943 & 0 \\
		$M_d=2$ & 0.046 & 0.428 & 0.442 & 1.199 & 0.850 & 0.944 & 0 \\
		$M_d=3$ & 0.069 & 0.528 & 0.439 & 1.191 & 0.852 & 0.945 & 4 \\
		$M_d=4$ & 0.100 & 0.600 & 0.435 & 1.181 & 0.855 & 0.947 & 7 \\
		$M_d=5$ & 0.139 & 0.657 & 0.433 & 1.172 & 0.859 & 0.950 & 24 \\
		\bottomrule
	\end{tabular}}
	\vspace{-2ex}
	\caption{FPR and assignment quality after detection using different settings of $M_d$. A higher value of $M_d$ offers a better protection against large colluding parties (see \cref{fig:tpr}), but also increases the detection FPR. Nevertheless, assignment quality is minimally impacted even with a high FPR since the majority of false positives have low rank and are unlikely to be assigned to begin with.}
  \label{tab:fpr}
  \vspace{-2ex}
\end{table*}

\subsection{Effectiveness of the Robust Assignment Algorithm}

We evaluate the robust assignment algorithm against successful attacks from \cref{sec:cheating}.

\noindent\textbf{What percentage of attacks is accurately detected?}
\cref{fig:tpr} shows the true positive rate (TPR) of detecting malicious reviewers as a function of collusion size, $M_a$ (on the $x$-axis), for different values of the hyperparameter $M_d$.
First, we measure the algorithm against all attacks that succeeded against the undefended scoring model (\emph{cf.} \cref{fig:prob}).
The results show that when $M_a \leq M_d$, the detection TPR is very close to 100\%, which implies \emph{almost all} malicious reviewers are removed in this case.
The TPR decreases as the size of the collusion, $M_a$ increases but still provides some protection even when $M_a > M_d$.
For instance, when $M_a = 5$ and $M_d = 4$ (darkest blue line), approximately 40\% of the successful attacks are detected.
%In the right plot, we evaluate against attackers that not only succeed in moving the reviewer into the candidate set, but also achieve a top-5 rank after attack. Detection TPR remains close to 100\% when $M_a \leq M_d$, but is much lower when $M_a > M_d$ due to the manipulated score being higher.
Increasing $M_d$ will protect against larger colluding parties at the cost of increasing the false positive rate (FPR), that is, the number of times in which an honest reviewer is mistaken for an adversary. A high FPR can negatively impact the quality of the assignments.

The degree of knowledge that we assume the attacker may possess far exceed that of typical reviewers. As a result, \cref{fig:tpr} may drastically underestimate the efficacy of our detection framework for practical applications. We further formulate a stronger \emph{colluding black-box attack} and evaluate against it in the appendix. Our results are very encouraging as it suggests that conference organizers can obtain robustness against more than 80\% of successful colluding black-box attacks with $M_a=10$ when applying our detection framework.

\begin{table*}[t!]
\small
	\centering
	\resizebox{\textwidth}{!}{%
	\begin{tabular}{c|cccc|ccccc}
		\toprule
		\bf Defense & \multicolumn{4}{c|}{\bf Assignment Quality} & \multicolumn{5}{c}{\bf Detection TPR} \\
	   & \bf Frac. of pos. & \bf Avg. bid score & \bf Avg. TPMS & \bf Avg. max. TPMS & $\mathbf{M_a = 1}$ & $\mathbf{M_a = 2}$ & $\mathbf{M_a = 3}$ & $\mathbf{M_a = 4}$ & $\mathbf{M_a = 5}$ \\
		\midrule
    TRIM $(L=10000)$ & 0.439 & 1.19 & 0.848 & 0.943 & 0.201 & 0.081 & 0.037 & 0.035 & 0.054 \\
    TRIM $(L=30000)$ & 0.219 & 0.439 & 0.816 & 0.917 & 0.986 & 0.966 & 0.942 & 0.924 & 0.919 \\
		\midrule
		\cref{alg:framework} $(M_d=1)$ & 0.443 & 1.201 & 0.849 & 0.943 & 1.000 & 0.077 & 0.000 & 0.000 & 0.000 \\
		\cref{alg:framework} $(M_d=5)$ & 0.433 & 1.172 & 0.859 & 0.950 & 1.000 & 1.000 & 1.000 & 1.000 & 1.000 \\
		\bottomrule
	\end{tabular}}
	\vspace{-2ex}
	\caption{Comparison of assignment quality and detection TPR against white-box attack between the TRIM robust regression algorithm and our robust assignment algorithm. See text for details.}
  \label{tab:robust_regression}
  \vspace{-3ex}
\end{table*}

\noindent\textbf{What is the quality of the final assignments?}
To study the effect of false positives from detection on the final paper assignments, we also evaluate assignment quality in terms of \emph{fraction of positive bids}, \emph{average bid score}, \emph{average TPMS}, and \emph{average maximum TPMS} (\emph{i.e.}, maximum TPMS score among assigned reviewers for each paper averaged over all papers).
Higher values of these metrics indicate a higher assignment quality.
The first row in \cref{tab:fpr} shows the assignment quality when using the NeurIPS-2014~\citep{lawrence2014nips} relevance scores. As expected, it over-emphasizes positive bids, which constitutes its inherent vulnerability. The second line shows the assignment quality when using only the TPMS score, which serves as a baseline for evaluating how much utility from bids is our robust assignment framework preserving. In contrast, using TPMS scores over-emphasizes average TPMS and average maximum TPMS.

The third line shows our assignment algorithm using the linear regression model without malicious reviewer detection ($M_d=0$). As it fills in the initially sparse bidding matrix, it has significantly more papers to choose from and yields assignments with fewer positive bids --- however the assignment quality is \emph{substantially higher} in terms of TPMS metrics compared to when using NeurIPS-2014 scores. The regression model offers a practical trade-off between relying on bids that reflect reviewer preference and relying on factors related to expertise (such as TPMS).

The remaining rows report results for the robust assignment algorithm with increasing values of $M_d$.
As expected, detection FPR increases as $M_d$ increases, but only has a limited effect on the assignment quality metrics.
The main reason for this is that most false positives are low-ranked reviewers, who are unlikely to be assigned the paper even if they were not excluded from the candidate set. Indeed, detection FPR is significantly lower for top-5 reviewers (second column) compared to that of top-50 reviewers (third column).
Overall, our results show that the assignment quality is hardly impacted by the detection mechanism.

We observed that a small number of papers were not assigned sufficient reviewers because the detection removed too many reviewers from the set of candidate reviewers for those papers.
We report this number in the last column (\# of under-reviewed) of \cref{tab:fpr}.
Although this is certainly a shortcoming of the robust assignment algorithm, the number of papers with insufficient candidates is small enough that it is still practical for conference organizers to assign them manually.

\noindent\textbf{Comparison with robust regression.} One effective defense against label-poisoning attacks for linear regression is the TRIM algorithm~\citep{jagielski2018manipulating}, which fits the model on a subset of the points that incur the least loss. The algorithm assumes that $L$ out of the $mn$ training points are poisoned and optimize:
\begin{align*}
\min_{\bw, \mathcal{I}} \quad &\| X^{\mathcal{I}} \bw - \by^{\mathcal{I}} \|_2^2 + \lambda \|\bw\|_2^2 \\
\text{s.t.} \quad &\mathcal{I} \subseteq \{1,\ldots,mn\}, |\mathcal{I}| = mn - L,
\end{align*}
where $X^{\mathcal{I}}, \by^{\mathcal{I}}$ denote the subset of $mn - L$ training data points selected by the index set $\mathcal{I}$. We apply TRIM to identify the $L$ poisoned pairs $(r,p)$ and remove them from the assignment candidate set. We then proceed to assign the remaining $mn - L$ pairs using \cref{eq:relevance}.

\cref{tab:robust_regression} shows the comparison between TRIM and our robust assignment algorithm in terms of assignment quality and detection TPR. The first and third rows correspond to the TRIM algorithm and \cref{alg:framework} that achieve a comparable assignment quality. Both methods fail to detect colluding attacks with $M_a > 1$, but \cref{alg:framework} is drastically more effective when $M_a = 1$. The second and fourth rows compare settings of TRIM and \cref{alg:framework} that achieve a similar detection TPR. Indeed, both have close to $100\%$ detection rate for $M_a=1,\ldots,5$. However, the assignment quality for TRIM is much worse, with all quality metrics being lower than using TPMS score alone (\emph{cf.} row 2 in \cref{tab:fpr}). Note that TRIM requires a drastic \emph{overestimate} of the number of poisoned data $(L=30,000)$ in order to detect most attack instances, which means that many benign training samples are being misidentified as malicious.

\noindent\textbf{Running time.}
As described in \cref{sec:detection}, our detection algorithm has a computational complexity of $O(d^2 + mnd + m \log M_d)$ for each reviewer-paper pair. In practice, pairs belonging to the same paper can be processed in a batch to re-use intermediate computation, which amounts to an average of 26 seconds per paper. This process can be easily parallelized across papers for efficiency.
% OLD TEXT: Paste back in emergency
% The first row in the table shows the assignment quality of our assignment algorithm without malicious reviewer detection ($M_d=0$); the remaining rows report results for the robust assignment algorithm with increasing values of $M_d$.
% As expected, the FPR of the detection increases as $M_d$ increases.
% However, these increases in FPR have only a limited effect on the four metrics of assignment quality that we consider.
% The main reason for this is that most false positives are low-ranked reviewers, which were unlikely to be assigned the paper even if they were not excluded from the candidate set. Indeed, detection FPR is significantly lower for top-5 reviewers (second column) compared to that of top-50 reviewers (third column).
% Overall, our results show that the assignment quality is hardly negatively affected by the mechanism that identifies and removes potentially adversarial reviewers.

% In our experiments, we did observe that a small number of papers were not assigned sufficient reviewers because the detection removed too many reviewers from the set of candidate reviewers for those papers.
% We report this number in the last column of \cref{tab:fpr}.
% Although this is certainly a shortcoming of the robust assignment algorithm, the number of papers with insufficient assignments is small enough that it is still practical for conference organizers to assign them manually.

%!TEX root=../icml_main.tex
\section{Related Work}\label{sec:related}
Our work fits in a larger body of work on automatic paper assignment systems, which includes studies on the design of relevance scoring functions~\citep{dumais1992automating, mimno2007expertise, rodriguez2008algorithm, liu2014robust} and appropriate quality metrics~\citep{goldsmith2007ai, tang2012optimization}. These studies have contributed to the development of conference management platforms such as EasyChair, HotCRP, and CMT that support most major computer science conferences.

Despite advances in automatic paper assignment, \cite{Rennie16} highlights shortcomings of peer-review systems owing to issues such as prejudices, misunderstandings, and corruption, all of which serve to make the system inefficient. For instance, the standard objective for assignment (say, \cref{eq:assignment_prob}) seeks to maximize the total relevance of assigned reviewers for the entire conference, which may be unfair to papers from under-represented areas. This has led to efforts that design objective functions and constraints to promote fairness in the assignment process for all submitted papers~\citep{garg2010assigning, long2013good, stelmakh2018peerreview4all, kobren2019paper}.

Furthermore, the assignment problem faces the additional challenge of coping with the implicit bias of reviewers~\citep{stelmakh2019testing}. This issue is particularly prevalent when authors of competing submissions participate in the review process, as they have an incentive to provide negative reviews in order to increase the chance of their own paper being accepted~\citep{anderson2007perverse, thurner2011peer}. In order to alleviate this problem, recent studies have devised assignment algorithms that promote impartiality in reviewers~\citep{aziz2016strategyproof, xu2018strategyproof}. We contribute to this line of work by identifying and removing reviewers who adversarially alter their bids to be assigned papers for which they have adverse incentives.

More recently, \citet{jecmen2020mitigating} studied the bid manipulation problem and considered an orthogonal approach to defending against it. Their method focuses on probabilistic assignment and upper limits the assignment probability for any paper-reviewer pair. As a result, the success rate of a bid manipulation attack is reduced. In contrast, our work seeks to limit the \emph{disproportional influence} of malicious bids rather than uniformly across all paper-reviewer pairs, and further considers the influence of colluding attackers on the assignment system.

%!TEX root=../aistats_main.tex
\section{Conclusion}
\label{sec:conclusion}
This study demonstrates some of the risks of paper bidding mechanisms that are commonly utilized in computer-science conferences to assign reviewers to paper submissions.
Specifically, we show that bid manipulation attacks may allow adversarial reviewers to review papers written by friends or rivals, even when these papers are outside of their area of expertise.
We developed a novel paper assignment system that is robust against such bid manipulation attacks, even in settings when multiple adversaries collude and have in-depth knowledge about the assignment system.
Our experiments on a synthetic but realistic dataset of conference papers demonstrate that our assignment system is, indeed, robust against such powerful attacks.
At the same time, our system still produces high-quality paper assignments for honest reviewers.
%Despite this increased robustness, our approach yields substantially higher quality paper assignments, as it imputes the bidding matrix, yielding higher assignment flexibility.
Our assignment algorithm is computationally efficient, easy to implement, and should be straightforward to incorporate into modern conference management systems.
%As a result, it should be easy to incorporate in existing conference management systems to deter reviewers from abusing the bidding mechanism.
We hope that our study contributes to a growing body of work aimed at developing techniques that can help improve the fairness, objectivity, and quality of the scientific peer-review process at scale.

\vspace{3ex}
{\large \textbf{Acknowledgements}}

This research is supported by grants from the National Science Foundation NSF
(III-1618134, III- 1526012, IIS-1149882, IIS-1724282, and TRIPODS-1740822, OAC-1934714),
the Bill and Melinda Gates Foundation, and the Cornell Center for Materials
Research with funding from the NSF MRSEC program (DMR-1719875), and SAP America.

\newpage
\bibliography{citations}

\begin{thebibliography}{35}
\providecommand{\natexlab}[1]{#1}
\providecommand{\url}[1]{\texttt{#1}}
\expandafter\ifx\csname urlstyle\endcsname\relax
  \providecommand{\doi}[1]{doi: #1}\else
  \providecommand{\doi}{doi: \begingroup \urlstyle{rm}\Url}\fi

\bibitem[Ammar et~al.(2018)Ammar, Groeneveld, Bhagavatula, Beltagy, Crawford,
  Downey, Dunkelberger, Elgohary, Feldman, Ha, et~al.]{ammar2018construction}
Ammar, W., Groeneveld, D., Bhagavatula, C., Beltagy, I., Crawford, M., Downey,
  D., Dunkelberger, J., Elgohary, A., Feldman, S., Ha, V., et~al.
\newblock Construction of the literature graph in semantic scholar.
\newblock \emph{arXiv preprint arXiv:1805.02262}, 2018.

\bibitem[Anderson et~al.(2007)Anderson, Ronning, De~Vries, and
  Martinson]{anderson2007perverse}
Anderson, M.~S., Ronning, E.~A., De~Vries, R., and Martinson, B.~C.
\newblock The perverse effects of competition on scientists’ work and
  relationships.
\newblock \emph{Science and engineering ethics}, 13\penalty0 (4):\penalty0
  437--461, 2007.

\bibitem[Aziz et~al.(2016)Aziz, Lev, Mattei, Rosenschein, and
  Walsh]{aziz2016strategyproof}
Aziz, H., Lev, O., Mattei, N., Rosenschein, J.~S., and Walsh, T.
\newblock Strategyproof peer selection: Mechanisms, analyses, and experiments.
\newblock In \emph{Thirtieth AAAI Conference on Artificial Intelligence}, 2016.

\bibitem[Bhatia et~al.(2015)Bhatia, Jain, and Kar]{bhatia2015robust}
Bhatia, K., Jain, P., and Kar, P.
\newblock Robust regression via hard thresholding.
\newblock In \emph{Advances in Neural Information Processing Systems}, pp.\
  721--729, 2015.

\bibitem[Biggio et~al.(2012)Biggio, Nelson, and Laskov]{biggio2012poisoning}
Biggio, B., Nelson, B., and Laskov, P.
\newblock Poisoning attacks against support vector machines.
\newblock \emph{arXiv preprint arXiv:1206.6389}, 2012.

\bibitem[Charlin \& Zemel(2013)Charlin and Zemel]{charlin2013toronto}
Charlin, L. and Zemel, R.
\newblock The toronto paper matching system: an automated paper-reviewer
  assignment system.
\newblock In \emph{ICML}, 2013.

\bibitem[Chen et~al.(2013)Chen, Caramanis, and Mannor]{chen2013robust}
Chen, Y., Caramanis, C., and Mannor, S.
\newblock Robust sparse regression under adversarial corruption.
\newblock In \emph{International Conference on Machine Learning}, pp.\
  774--782, 2013.

\bibitem[Cormen et~al.(2009)Cormen, Leiserson, Rivest, and Stein]{clrs2009}
Cormen, T.~H., Leiserson, C.~E., Rivest, R.~L., and Stein, C.
\newblock \emph{Introduction to Algorithms, Third Edition}.
\newblock The MIT Press, 3rd edition, 2009.
\newblock ISBN 0262033844.

\bibitem[Dean \& Henzinger(1999)Dean and Henzinger]{dean1999finding}
Dean, J. and Henzinger, M.~R.
\newblock Finding related pages in the world wide web.
\newblock \emph{Computer networks}, 31\penalty0 (11-16):\penalty0 1467--1479,
  1999.

\bibitem[Dumais \& Nielsen(1992)Dumais and Nielsen]{dumais1992automating}
Dumais, S.~T. and Nielsen, J.
\newblock Automating the assignment of submitted manuscripts to reviewers.
\newblock In \emph{Proceedings of the 15th annual international ACM SIGIR
  conference on Research and development in information retrieval}, pp.\
  233--244, 1992.

\bibitem[Garg et~al.(2010)Garg, Kavitha, Kumar, Mehlhorn, and
  Mestre]{garg2010assigning}
Garg, N., Kavitha, T., Kumar, A., Mehlhorn, K., and Mestre, J.
\newblock Assigning papers to referees.
\newblock \emph{Algorithmica}, 58\penalty0 (1):\penalty0 119--136, 2010.

\bibitem[Goldsmith \& Sloan(2007)Goldsmith and Sloan]{goldsmith2007ai}
Goldsmith, J. and Sloan, R.~H.
\newblock The ai conference paper assignment problem.
\newblock In \emph{Proc. AAAI Workshop on Preference Handling for Artificial
  Intelligence, Vancouver}, pp.\  53--57, 2007.

\bibitem[Hartvigsen et~al.(1999)Hartvigsen, Wei, and
  Czuchlewski]{hartvigsen1999conference}
Hartvigsen, D., Wei, J.~C., and Czuchlewski, R.
\newblock The conference paper-reviewer assignment problem.
\newblock \emph{Decision Sciences}, 30\penalty0 (3):\penalty0 865--876, 1999.

\bibitem[Jagielski et~al.(2018)Jagielski, Oprea, Biggio, Liu, Nita-Rotaru, and
  Li]{jagielski2018manipulating}
Jagielski, M., Oprea, A., Biggio, B., Liu, C., Nita-Rotaru, C., and Li, B.
\newblock Manipulating machine learning: Poisoning attacks and countermeasures
  for regression learning.
\newblock In \emph{2018 IEEE Symposium on Security and Privacy (SP)}, pp.\
  19--35. IEEE, 2018.

\bibitem[Jecmen et~al.(2020)Jecmen, Zhang, Liu, Shah, Conitzer, and
  Fang]{jecmen2020mitigating}
Jecmen, S., Zhang, H., Liu, R., Shah, N.~B., Conitzer, V., and Fang, F.
\newblock Mitigating manipulation in peer review via randomized reviewer
  assignments.
\newblock \emph{arXiv preprint arXiv:2006.16437}, 2020.

\bibitem[Kobren et~al.(2019)Kobren, Saha, and McCallum]{kobren2019paper}
Kobren, A., Saha, B., and McCallum, A.
\newblock Paper matching with local fairness constraints.
\newblock In \emph{Proceedings of the 25th ACM SIGKDD International Conference
  on Knowledge Discovery \& Data Mining}, pp.\  1247--1257, 2019.

\bibitem[Koh et~al.(2018)Koh, Steinhardt, and Liang]{koh2018stronger}
Koh, P.~W., Steinhardt, J., and Liang, P.
\newblock Stronger data poisoning attacks break data sanitization defenses.
\newblock \emph{arXiv preprint arXiv:1811.00741}, 2018.

\bibitem[Kuhn(1955)]{kuhn1955hungarian}
Kuhn, H.~W.
\newblock The hungarian method for the assignment problem.
\newblock \emph{Naval research logistics quarterly}, 2\penalty0 (1-2):\penalty0
  83--97, 1955.

\bibitem[Lawrence(2014)]{lawrence2014nips}
Lawrence, N.
\newblock Paper allocation for nips, 2014.
\newblock
  \url{https://inverseprobability.com/2014/06/28/paper-allocation-for-nips}.
  [Online; accessed on 2020-10-02].

\bibitem[Liu et~al.(2014)Liu, Suel, and Memon]{liu2014robust}
Liu, X., Suel, T., and Memon, N.
\newblock A robust model for paper reviewer assignment.
\newblock In \emph{Proceedings of the 8th ACM Conference on Recommender
  systems}, pp.\  25--32, 2014.

\bibitem[Long et~al.(2013)Long, Wong, Peng, and Ye]{long2013good}
Long, C., Wong, R. C.-W., Peng, Y., and Ye, L.
\newblock On good and fair paper-reviewer assignment.
\newblock In \emph{2013 IEEE 13th International Conference on Data Mining},
  pp.\  1145--1150. IEEE, 2013.

\bibitem[Mei \& Zhu(2015)Mei and Zhu]{mei2015using}
Mei, S. and Zhu, X.
\newblock Using machine teaching to identify optimal training-set attacks on
  machine learners.
\newblock In \emph{Proceedings of the AAAI Conference on Artificial
  Intelligence}, volume~29, 2015.

\bibitem[Mimno \& McCallum(2007)Mimno and McCallum]{mimno2007expertise}
Mimno, D. and McCallum, A.
\newblock Expertise modeling for matching papers with reviewers.
\newblock In \emph{Proceedings of the 13th ACM SIGKDD international conference
  on Knowledge discovery and data mining}, pp.\  500--509, 2007.

\bibitem[Rennie(2016)]{Rennie16}
Rennie, D.
\newblock Let's make peer review scientific.
\newblock \emph{Nature}, 2016.

\bibitem[Rodriguez \& Bollen(2008)Rodriguez and Bollen]{rodriguez2008algorithm}
Rodriguez, M.~A. and Bollen, J.
\newblock An algorithm to determine peer-reviewers.
\newblock In \emph{Proceedings of the 17th ACM conference on Information and
  knowledge management}, pp.\  319--328, 2008.

\bibitem[Shah et~al.(2018)Shah, Tabibian, Muandet, Guyon, and
  Von~Luxburg]{shah2018design}
Shah, N.~B., Tabibian, B., Muandet, K., Guyon, I., and Von~Luxburg, U.
\newblock Design and analysis of the nips 2016 review process.
\newblock \emph{The Journal of Machine Learning Research}, 19\penalty0
  (1):\penalty0 1913--1946, 2018.

\bibitem[Stelmakh et~al.(2018)Stelmakh, Shah, and
  Singh]{stelmakh2018peerreview4all}
Stelmakh, I., Shah, N.~B., and Singh, A.
\newblock Peerreview4all: Fair and accurate reviewer assignment in peer review.
\newblock \emph{arXiv preprint arXiv:1806.06237}, 2018.

\bibitem[Stelmakh et~al.(2019)Stelmakh, Shah, and Singh]{stelmakh2019testing}
Stelmakh, I., Shah, N., and Singh, A.
\newblock On testing for biases in peer review.
\newblock In \emph{Advances in Neural Information Processing Systems}, pp.\
  5287--5297, 2019.

\bibitem[Stent \& Ji(2018)Stent and Ji]{stent2018naacl}
Stent, A. and Ji, H.
\newblock A review of reviewer assignment methods, 2018.
\newblock
  \url{https://naacl2018.wordpress.com/2018/01/28/a-review-of-reviewer-assignment-methods}.
  [Online; accessed on 2020-10-02].

\bibitem[Tang et~al.(2012)Tang, Tang, Lei, Tan, Gao, and
  Li]{tang2012optimization}
Tang, W., Tang, J., Lei, T., Tan, C., Gao, B., and Li, T.
\newblock On optimization of expertise matching with various constraints.
\newblock \emph{Neurocomputing}, 76\penalty0 (1):\penalty0 71--83, 2012.

\bibitem[Thurner \& Hanel(2011)Thurner and Hanel]{thurner2011peer}
Thurner, S. and Hanel, R.
\newblock Peer-review in a world with rational scientists: Toward selection of
  the average.
\newblock \emph{The European Physical Journal B}, 84\penalty0 (4):\penalty0
  707--711, 2011.

\bibitem[Vijaykumar(2020)]{vijaykumar2020potential}
Vijaykumar, T.~N.
\newblock Potential organized fraud in acm/ieee computer architecture
  conferences, 2020.
\newblock
  \url{https://medium.com/@tnvijayk/potential-organized-fraud-in-acm-ieee-computer-architecture-conferences-ccd61169370d}.
  [Online; accessed on 2020-10-13].

\bibitem[Weinberger et~al.(2009)Weinberger, Dasgupta, Langford, Smola, and
  Attenberg]{weinberger2009feature}
Weinberger, K., Dasgupta, A., Langford, J., Smola, A., and Attenberg, J.
\newblock Feature hashing for large scale multitask learning.
\newblock In \emph{Proceedings of the 26th annual international conference on
  machine learning}, pp.\  1113--1120, 2009.

\bibitem[Xiao et~al.(2015)Xiao, Biggio, Nelson, Xiao, Eckert, and
  Roli]{xiao2015support}
Xiao, H., Biggio, B., Nelson, B., Xiao, H., Eckert, C., and Roli, F.
\newblock Support vector machines under adversarial label contamination.
\newblock \emph{Neurocomputing}, 160:\penalty0 53--62, 2015.

\bibitem[Xu et~al.(2018)Xu, Zhao, Shi, and Shah]{xu2018strategyproof}
Xu, Y., Zhao, H., Shi, X., and Shah, N.~B.
\newblock On strategyproof conference peer review.
\newblock \emph{arXiv preprint arXiv:1806.06266}, 2018.

\end{thebibliography}
\bibliographystyle{icml2021}

\newpage
\onecolumn
\appendix
%!TEX root=../icml_main.tex

\setcounter{equation}{0}
\setcounter{figure}{0}
\setcounter{table}{0}
\renewcommand{\theequation}{S\arabic{equation}}
\renewcommand{\thefigure}{S\arabic{figure}}
\renewcommand{\thetable}{S\arabic{table}}
\noindent\makebox[\linewidth]{\rule{\linewidth}{3.5pt}}
\begin{center}
	\bf{\Large Supplementary Material: Making Paper Reviewing Robust to Bid Manipulation Attacks}
\end{center}
\noindent\makebox[\linewidth]{\rule{\linewidth}{1pt}}

\section{Dataset Construction}
\label{sec:dataset}

In this section, we describe how we subsampled data from the Semantic Scholar Open Research Corpus (S2ORC)~\citep{ammar2018construction}, extracted reviewer/paper features such as subject area and TPMS, and simulated bids using citation. Our data is publicly released\footnote{\url{https://drive.google.com/drive/folders/1khI9kaPy_8F0GtAzwR-48Jc3rsQmBhfe?usp=sharing}} for reproducibility and to facilitate future research.

\subsection{Conference Simulation}
\label{sec:simulation}

The goal of our dataset is to simulate a NeurIPS-like conference environment, where the organizers assign reviewers to papers based on expertise and interest. We first retrieve the collection of 6956 papers from S2ORC that are published in ML/AI/CV/NLP venues between the years 2014-2015, which includes the following conferences: AAAI, AISTATS, ACL, COLT, CVPR, ECCV, EMNLP, ICCV, ICLR, ICML, IJCAI, NeurIPS, and UAI. We believe the diversity of subject areas represented by the above conferences is an accurate reflection of typical ML/AL conferences in recent years. We will refer to this collection of papers as the \emph{corpus}.

\paragraph{Subject areas.} Most conferences require authors to indicate primary and secondary subject areas for their submitted papers. However, the S2ORC only contains a \emph{field of study} attribute for most of the retrieved papers in the corpus, which is often the broad category of \emph{computer science}. To identify the suitable fine-grained subjects for each paper, we adopt an unsupervised learning approach of clustering the papers by relatedness and treating each discovered cluster as a subject area.

Similarity is defined in terms of co-citations -- a common signal used in information retrieval for discovering related documents~\citep{dean1999finding}. For a paper $p$, let $N(p)$ denote the union of in-citations and out-citations for $p$. The similarity between two papers $p,q$ is defined as
\begin{equation}
\sigma(p,q) = \frac{|N(p) \cap N(q)|}{\sqrt{|N(p)|} \cdot \sqrt{|N(q)|}},
\label{eq:paper_sim}
\end{equation}
which is the cosine similarity in document retrieval. We perform agglomerative clustering using average linkage\footnote{\url{https://scikit-learn.org/stable/modules/clustering.html\#hierarchical-clustering}} to reduce the set of papers to 1000 clusters. After removing small cluster (less than 5 papers), we obtain 368 clusters to serve as subject areas. \cref{tab:subject_areas} shows a few sample clusters along with papers contained in the cluster. Most of the discovered clusters are highly coherent with members sharing keywords in their titles despite the definition of similarity depending \emph{entirely} on co-citations.

To populate the list of subject areas for a given paper $p$, we first compute its subject relatedness to a cluster $C$ by:
\begin{equation}
\sigma(p, C) = \frac{1}{|C|} \sum_{q \in C} \sigma(p, q).
\label{eq:subject_sim}
\end{equation}
Given the set of clusters representing subject areas, we identify the top-5 clusters according to $\sigma(p,C)$ to be the list of subject areas for the paper $p$, denoted $\text{subj}(p)$.

\begin{table*}[ht!]
	\centering
	\resizebox{\textwidth}{!}{%
	\begin{tabular}{l|l}
		\toprule
		{\bf Subject Area} & {\bf Papers} \\
		\midrule
		Multi-task learning & \tabincell{l}{Encoding Tree Sparsity in Multi-Task Learning: A Probabilistic Framework\\ Multi-Task Learning and Algorithmic Stability\\ Exploiting Task-Feature Co-Clusters in Multi-Task Learning\\ Efficient Output Kernel Learning for Multiple Tasks\\ Learning Multiple Tasks with Multilinear Relationship Networks\\ \emph{Etc.}} \\
		\midrule
    	Video segmentation & \tabincell{l}{Efficient Video Segmentation Using Parametric Graph Partitioning\\ Video Segmentation with Just a Few Strokes\\ Co-localization in Real-World Images\\ Semantic Single Video Segmentation with Robust Graph Representation\\ PatchCut: Data-driven object segmentation via local shape transfer\\ \emph{Etc.}} \\
    	\midrule
		Topic modeling &
		\tabincell{l}{On Conceptual Labeling of a Bag of Words\\ Topic Modeling with Document Relative Similarities\\ Divide-and-Conquer Learning by Anchoring a Conical Hull\\ Spectral Methods for Supervised Topic Models\\ Model Selection for Topic Models via Spectral Decomposition\\ \emph{Etc.}} \\
		\bottomrule
		Feature selection &
		\tabincell{l}{Embedded Unsupervised Feature Selection\\ Feature Selection at the Discrete Limit\\ Bayes Optimal Feature Selection for Supervised Learning with General Performance Measures\\ Reconsidering Mutual Information Based Feature Selection: A Statistical Significance View\\ Unsupervised Simultaneous Orthogonal basis Clustering Feature Selection\\ \emph{Etc.}} \\
		\bottomrule
	\end{tabular}
	}
	\caption{Sample subject areas and paper titles of cluster members.}
	\vspace{1ex}
  \label{tab:subject_areas}
\end{table*}

\paragraph{Reviewers.} The S2ORC dataset contains entries of authors along with their list of published papers. We utilize this information to simulate reviewers by collecting the set of authors who has cited at least one paper from the corpus. The total number of retrieved authors is 234,598. Because the vast majority of retrieved authors are very loosely related to the field of ML/AI, they would not be suitable reviewer candidates for a real ML/AI conference. Therefore, we retain only authors who have cited at least 15 papers from the corpus to serve as reviewers. We also remove authors who cited more than 50 papers from the corpus, since these reviewers represent senior researchers that would typically serve as area chairs. The number of remaining reviewers is $5,914$.

Most conferences also solicit self-reported subject areas from reviewers. We simulate this attribute by leveraging the clusters discovered through co-citation. For each subject area $C$, we count the number of times $C$ appeared in $\text{subj}(p)$ for each of the papers $p$ that the reviewer $r$ has cited. The 5 most frequently appearing clusters (ties are broken randomly) serve as the reviewer's subject areas, denoted $\text{subj}(r)$.

\paragraph{TPMS score.} The TPMS score~\citep{charlin2013toronto} is computed by measuring the similarity between a reviewer's profile -- represented by a set of papers that the reviewer uploads -- and a target paper. We simulate this score using the language model-based approach from the original TPMS paper, which we detail below for completeness. For a reviewer $r$, let $A_r$ denote the bag-of-words representation for the set of papers that the reviewer has authored. More specifically, we collect the abstracts of the papers that $r$ has authored, remove all stop words, and pool the remaining words together into $A_r$ as a multi-set. Similarly, let $A_p$ denote the bag-of-words representation for the abstract of a paper $p$. The simulated TPMS is computed as:
\begin{equation}
\text{TPMS}_{r,p} = \sum_{w \in A_p} \log f_{rw},
\label{eq:tpms}
\end{equation}
where $f_{rw}$ is the Dirichlet-smoothed normalized frequency of the word $w$ in $A_r$. Let $D$ denote the bag-of-words representation for the entire corpus of (abstracts of) papers, and let $D(w)$ (resp. $A_r(w)$) denote the occurrences of $w$ in the corpus (resp. $A_r$). Then $$f_{rw} := \left( \frac{|A_r|}{|A_r| + \beta} \right) \frac{|A_r(w)|}{|A_r|} + \left( \frac{\beta}{|A_r| + \beta} \right) \frac{|D(w)|}{|D|},$$ where $\beta$ is a smoothing factor. We set $\beta = 1000$ in our experiment. The obtained scores are normalized per paper between 0 and 1.

\subsection{Simulating Bids}
\label{sec:bids}

The most challenging aspect of our simulation is the bids. At first, it may seem natural to simulate bids using citations, since it is a proxy of interest and can be easily obtained from the S2ORC dataset. However, we have observed that bids are heavily skewed towards a few very influential papers, while the distribution of bids is much more uniform across all papers. To overcome this issue, we instead model a reviewer's bidding behavior based on the following assumptions:
\begin{enumerate}[1.]
\item A reviewer will only bid on papers from subject areas that he/she is familiar with.
\item Given two papers from the same subject area, a reviewer favors bidding on a paper whose title/abstract is a better match with the reviewer's profile.
\end{enumerate}
We define several scores that reflect the above aspects and combine them to obtain the final bids. In practice, reviewers will often also rely on TPMS to sort the papers to bid on. However, since our simulated TPMS depends entirely on the abstract, we omit TPMS in our bidding model. Nevertheless, we have observed empirically that TPMS is highly correlated with the bids that we obtain.

\paragraph{Subject score.} We leverage citation to reflect the degree of interest in the subject of a paper. Let $\text{icf}(q)$ denote the \emph{inverse citation frequency} (ICF) of a paper $q$ in the corpus: $$\text{icf}(q) = \log \frac{\text{\# total in-citations in the corpus}}{\text{\# in-citations for }q}.$$ The purpose of the ICF is to down-weight commonly cited papers to avoid overcrowding of bids. Denote by $C^*(q)$ the top cluster that $q$ belongs to according to \cref{eq:subject_sim}. The \emph{subject score} for a paper $p$ is defined as:
\begin{equation}
\text{subject-score}_{r,p} = \sum_{q : r \text{ cites } q} \frac{\text{icf}(q)}{|C^*(q)|} \mathds{1}\{p \in C^*(q)\}.
\label{eq:subject_score}
\end{equation}
In other words, for each paper $q$ that $r$ cites, we merge all papers from the same subject area of $q$, represented by $C^*(q)$, into the reviewer's pool. Each paper in $C^*(q)$ is weighted by the reciprocal of the cluster size and the ICF of $q$, and the subject score is the resulting sum after accumulating over all papers $q$ that the reviewer cites. Note that every paper within the same subject cluster has the \emph{exact same} subject score, which is non-zero only if the reviewer has bid on a paper within this subject area. This property reflects the assumption that a reviewer is only interested in papers from familiar subject areas, and is indifferent to different papers in the same subject absent of title/abstract information. To avoid overcrowding by frequently cited papers, we set $\text{subject-score}_{r,p} = 0$ for any paper $p$ that received over 1000 citations.

\begin{figure*}[t!]
	\includegraphics[width=1\linewidth]{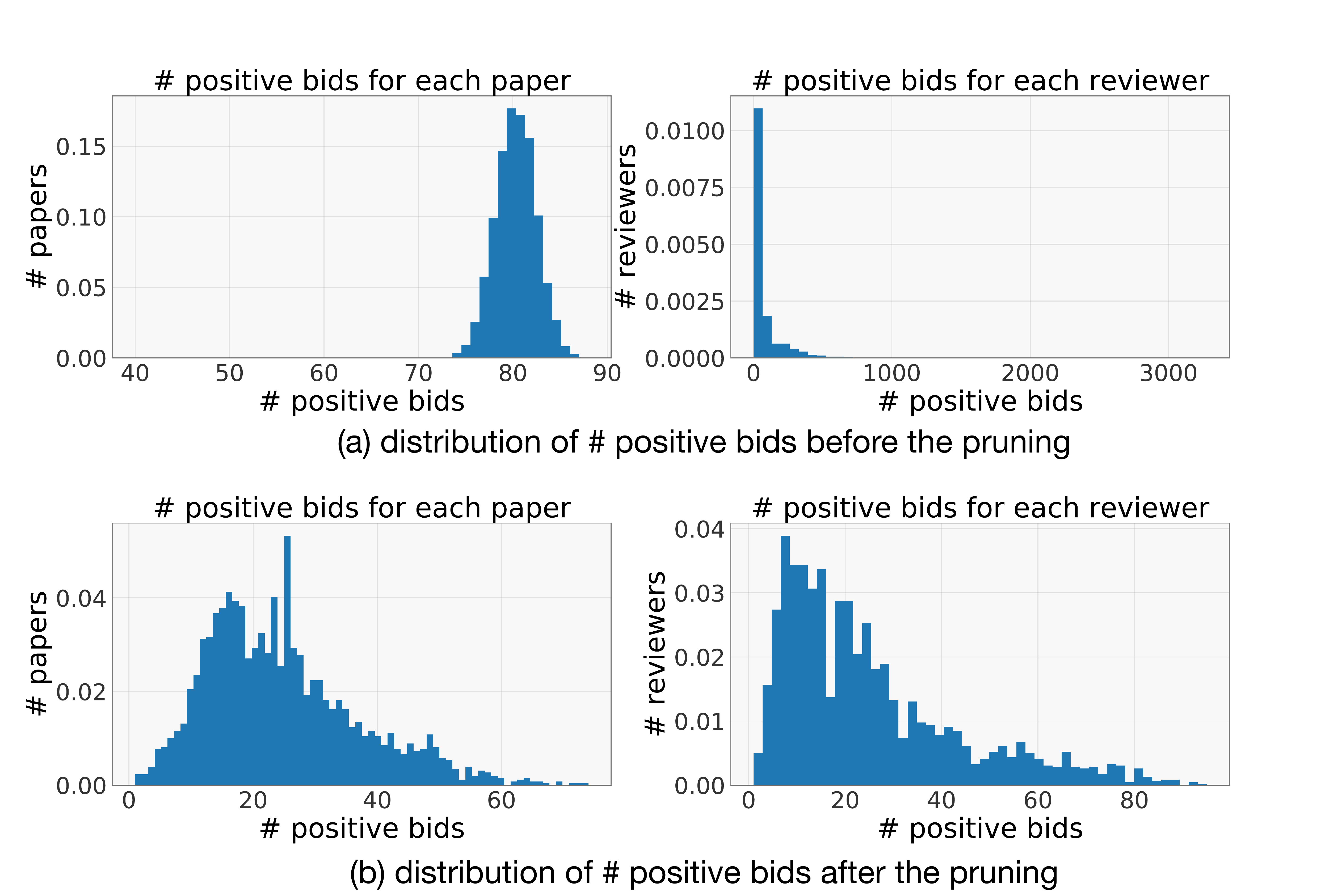}
	\caption{Distribution of the number of positive bids before and after subsampling.}
	\vspace{1ex}
	\label{fig:pos_dist}
\end{figure*}

\paragraph{Title/abstract score.} To measure the degree of title/abstract similarity between a reviewer and a paper, we compute the inner product between the TF-IDF vectors of the reviewer's and paper's title/abstract. Let $\text{idf}(w)$ denote the \emph{inverse document frequency} of a word $w$. For each reviewer $r$, let $\text{tf-idf}(r)$ denote the vector, indexed by words, such that $\text{tf-idf}(r)_w = (|A_r(w)| / |A_r|) \cdot \text{idf}(w)$ for each word $w$. Similarly, we can define the TF-IDF vector for a paper $p$, and the \emph{abstract score} between a pair $(r,p)$ is given by the inner product:
\begin{equation}
\text{abstract-score}_{r,p} = \text{tf-idf}(r) \cdot \text{tf-idf}(p).
\label{eq:abstract_score}
\end{equation}
We can define the \emph{title score} in an analogous manner based on the bag-of-words representation of titles instead of abstracts.

\paragraph{Bidding.} We simulate bids by combining the subject/title/abstract scores as follows. First, we define a \emph{total score}:
\begin{equation}
\text{total-score}_{r,p} = (\text{title-score}_{r,p} + \text{abstract-score}_{r,p}) \cdot \text{subject-score}_{r,p},
\end{equation}
which reflects the assumptions we made about a reviewer's bidding behavior, \emph{i.e.}, a higher total score reflects a higher reviewer interest in the paper. The total score gives us a ranking of papers in the corpus, denoted by $\text{rank}_r(p)$, for each paper $p$. To obtain the positive bids, we randomly retain high-ranked papers with a decaying probability: $$\Pr(r \text{ bids on } p) = 1 / (1 + \exp(\alpha \cdot (\text{rank}_r(p) - \mu) ),$$ where $\alpha$ and $\mu$ are hyperparameters that control the steepness of the drop in sampling probability for low-ranked papers, and the average number of papers that each reviewer bids on. We set $\alpha = 0.2$ and $\mu = 80$ in our experiment.

The quality of bids obtained from this sampling procedure is very reasonable. However, the majority of papers had very few bids (see \cref{fig:pos_dist}(a)) -- contrary to statistics observed in a real conference such as NeurIPS-2016 (see Figure 1 in~\cite{shah2018design}). To match the distribution of the number of bids per reviewer/paper to that of a real conference, we further subsample papers (resp. reviewers) to encourage selecting ones with more bids. The distribution of the number positive bids per reviewer/paper after subsampling is shown in \cref{fig:pos_dist}(b). Our finalized conference dataset contains $m=2483$ reviewers and $n=2446$ submitted papers -- a realistic balance of papers and reviewers for recent ML/AI conferences.

Finally, some conferences allow more fine-grained bids, such as \emph{in a pinch}, \emph{willing} and \emph{eager} for conferences managed using CMT. To simulate \emph{bid scores} that reflect the degree of interest, we quantize the total score of all positive bids into the discrete range $\{1,2,3\}$ based on the distribution of bid scores in a real conference: at a ratio of $8 : 53 : 39$ for the bids 1, 2 and 3.

\section{Features and Training}
\label{sec:features}

We provide details regarding feature extraction and model training in this section. To fully imitate a conference management environment, we extract relevant features from papers and reviewers that are obtainable in a realistic scenario, including: paper/reviewer subject area (5 areas for each), bag-of-words vector for paper title, and (simulated) TPMS. These features are further processed and concatenated as input to the linear regression model in \cref{sec:model}.

\cref{tab:feature_info} lists all the extracted features and their dimensions. Paper title (PT) is the vectorized count of words appearing in the paper's title, while paper subject area (PS), reviewer subject area (RS) and intersected subject area (IS) are categorical features represented using binary vectors. The first dimension for the TPMS vector (TV) is the TPMS score for the reviewer-paper pair. We also quantize the raw TPMS into 11 bins and use the bin index as well as the quantized scores, which results in the remaining 11 dimensions for the TPMS vector.

RS$\otimes$PS, RS$\otimes$PT, IS$\otimes$PT and IS$\otimes$TV are additional quadratic features that capture the interaction between feature pairs. The introduction of these quadratic features results in a very high-dimensional, albeit extremely sparse feature vector, and hence many dimensions could be collapsed without a significant impact to performance. We apply feature hashing~\cite{weinberger2009feature} to the quadratic features at a hash ratio of 0.01, which reduces the total feature dimensionality to $d=10,288$.

\begin{table*}[t!]
\small
	\centering
	\begin{tabular}{l|ccc}
		\toprule
		\bf Features & paper titles (PT) & paper subject area (PS) & reviewer subject area (RS)\\
		\midrule
		\bf \# of Dimensions & 930 & 368 & 368\\
		\toprule
		\bf Features  & intersected subject area (IS) & TPMS vector (TV) & RS$\otimes$PS \\
		\midrule
		\bf \# of Dimensions & 368 & 12 & 135424 \\
		\toprule
		\bf Features  & RS$\otimes$PT & IS$\otimes$PT & IS$\otimes$TV\\
		\midrule
		\bf \# of Dimensions & 342240 & 342240 & 4410 \\
		\bottomrule
	\end{tabular}
	\caption{Extracted features and their dimensionalities. See the text for details.}
	\vspace{1ex}
	\label{tab:feature_info}
\end{table*}

\begin{table*}[t!]
\small
	\centering
	\resizebox{\textwidth}{!}{%
	\begin{tabular}{ll|cccccccccc}
		\toprule
		~ & ~ &\bf  k=1 &\bf  k=2 &\bf  k=3 &\bf  k=4 &\bf  k=5 &\bf  k=6 &\bf  k=7 &\bf  k=8 &\bf  k=9 &\bf  k=10\\
		\midrule
    	\multirow{2}{*}{\bf AP@k per reviewer} & \bf train & 0.41 & 0.41 & 0.40 & 0.39 & 0.38 & 0.38 & 0.37 & 0.37 & 0.36 & 0.35\\
    	  & \bf test & 0.38 & 0.41 & 0.39 & 0.38 & 0.38 & 0.37 & 0.36 & 0.36 & 0.35 & 0.34\\
    	\midrule
		\multirow{2}{*}{\bf AP@k per paper} &\bf  train & 0.55 & 0.53 & 0.51 & 0.50 & 0.49 & 0.47 & 0.46 & 0.45 & 0.43 & 0.42 \\
		  & \bf test & 0.58 & 0.55 & 0.52 & 0.51 & 0.48 & 0.47 & 0.45 & 0.44 & 0.43 & 0.41 \\
		\bottomrule
	\end{tabular}
	}
	\caption{Average precision@k per reviewer/paper for the trained linear regressor.}
  	\label{tab:precision}
\end{table*}

\paragraph{Model performance.} To validate our linear regression model and the selected features, we test the average precision at k (AP@k) for the trained model on a train-test split. \cref{tab:precision} shows the AP@k per reviewer (P@k for finding papers relevant to a reviewer, averaged across all reviewers) and the AP@k per paper for the linear regressor. It is evident that both metrics are at an acceptable level for real world deployment, and the train-test gap is minimal, indicating that the model is able to generalize well beyond observed bids.

%For precision, we make the train-test split of our dataset by reviewers / papers and evaluate the average precision@k (how many positive bids in top $k$) among all reviewers / papers in training set and test set. \cref{tab:precision} shows the results for precision and as we can see precisions on both training set and test set are pretty sweet. Small gap between results for train set and test set shows the good generalization ability of our model.

\begin{table*}[t!]
	\centering
	\resizebox{\textwidth}{!}{%
	\begin{tabular}{l|l|c}
		\toprule
		{\bf Reviewer} & {\bf Assigned Papers} & {\bf Bid Scores} \\
		\midrule
    	Kavita Bala & \tabincell{l}{1. Learning Lightness from Human Judgement on Relative Reflectance\\
2. Simulating Makeup through Physics-Based Manipulation of Intrinsic Image Layers\\
3. Learning Ordinal Relationships for Mid-Level Vision\\
4. Automatically Discovering Local Visual Material Attributes\\
5. Recognize Complex Events from Static Images by Fusing Deep Channels\\
6. Learning a Discriminative Model for the Perception of Realism in Composite Images} & \tabincell{l}{3 \\ 3 \\ 3 \\ 0 \\ 0 \\ 0} \\
    	\midrule
    	Ryan P. Adams & \tabincell{l}{1. Stochastic Variational Inference for Hidden Markov Models\\
2. Parallel Markov Chain Monte Carlo for Pitman-Yor Mixture Models\\
3. Celeste: Variational Inference for a Generative Model of Astronomical Images\\
4. Measuring Sample Quality with Stein'S Method\\
5. Parallelizing MCMC with Random Partition Trees\\
6. Hamiltonian ABC} & \tabincell{l}{3 \\ 0 \\ 0 \\ 0 \\ 0 \\ 0} \\
		\midrule
    	Peter Stone & \tabincell{l}{1. Qualitative Planning with Quantitative Constraints for Online Learning of Robotic Behaviours\\
2. An Automated Measure of MDP Similarity for Transfer in Reinforcement Learning\\
3. On Convergence and Optimality of Best-Response Learning with Policy Types in Multiagent Systems\\
4. A Framework for Task Planning in Heterogeneous Multi Robot Systems Based on Robot Capabilities\\
5. A Strategy-Aware Technique for Learning Behaviors from Discrete Human Feedback\\
6. Stick-Breaking Policy Learning in Dec-Pomdps} & \tabincell{l}{3 \\ 3 \\ 3 \\ 3 \\ 0 \\ 0} \\
		\midrule
    	Yejin Choi & \tabincell{l}{1. Don'T Just Listen, Use Your Imagination: Leveraging Visual Common Sense for Non-Visual Tasks\\
2. Segment-Phrase Table for Semantic Segmentation, Visual Entailment and Paraphrasing\\
3. Refer-To-As Relations as Semantic Knowledge} & \tabincell{l}{3 \\ 0 \\ 0} \\
		\midrule
    	Emma Brunskill & \tabincell{l}{1. Policy Evaluation Using the $\Omega$-Return\\
2. Towards More Practical Reinforcement Learning\\
3. High Confidence Policy Improvement\\
4. Sample Efficient Reinforcement Learning With Gaussian Processes\\
5. Policy Tree: Adaptive Representation for Policy Gradient\\
6. Abstraction Selection in Model-Based Reinforcement Learning} & \tabincell{l}{3 \\ 3 \\ 3 \\ 3 \\ 3 \\ 0} \\
		\midrule
    	Elad Hazan & \tabincell{l}{1. Online Linear Optimization via Smoothing\\
2. Online Learning for Adversaries with Memory: Price of Past Mistakes\\
3. Hierarchies of Relaxations for Online Prediction Problems with Evolving Constraints\\
4. Hard-Margin Active Linear Regression\\
5. Online Gradient Boosting\\
6. Robust Multi-Objective Learning With Mentor Feedback} & \tabincell{l}{3 \\ 0 \\ 0 \\ 0 \\ 0 \\ 0} \\

		\bottomrule
	\end{tabular}
	}
	\caption{Assigned papers for six representative reviewers.}
	\vspace{1ex}
  \label{tab:case_study}
\end{table*}

\begin{figure}[t!]
\centering
\includegraphics[width=0.5\linewidth]{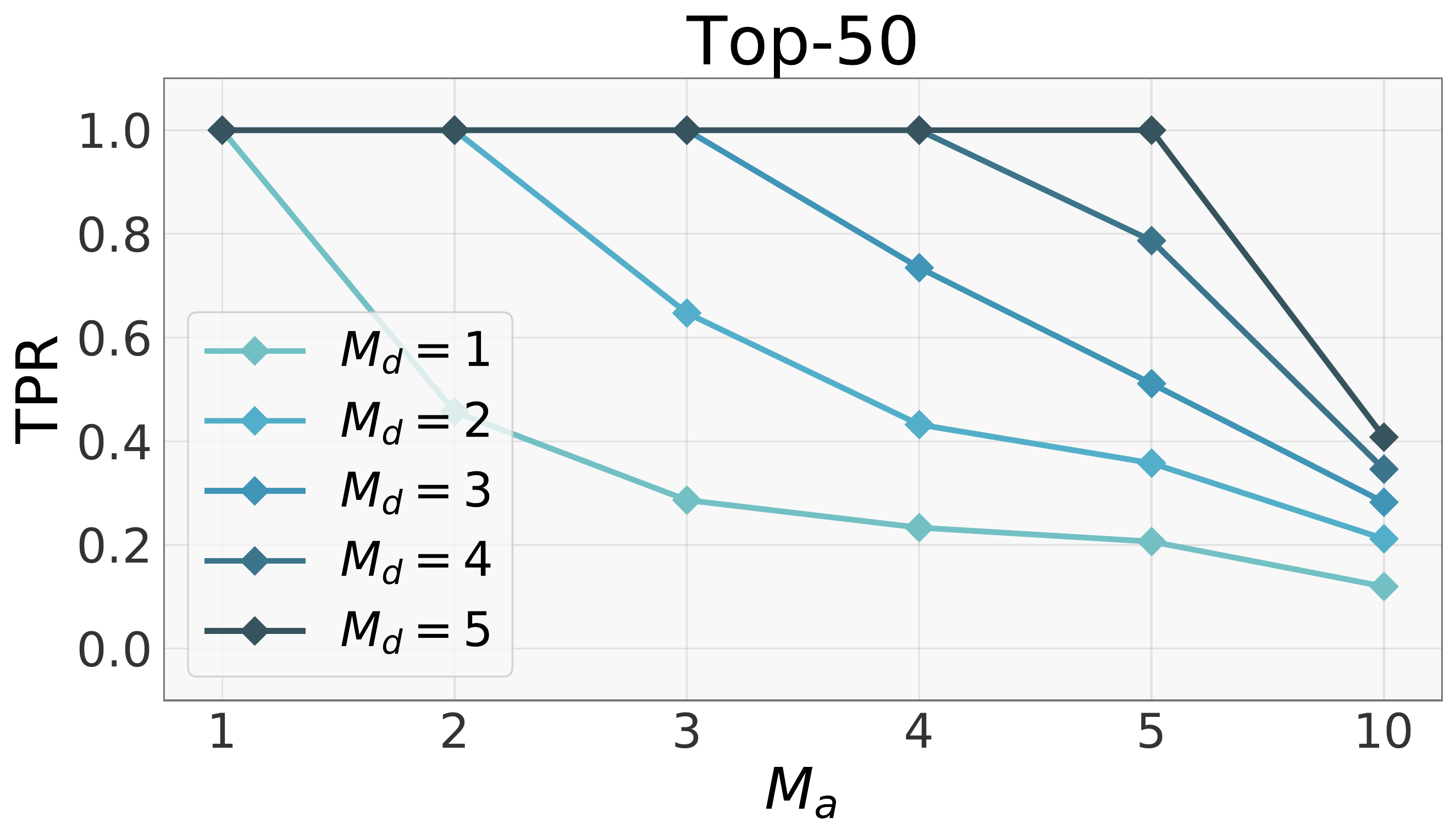}
\vspace{-2ex}
\caption{TPR for detecting \emph{colluding white-box attacks} that succeed in achieving top-50 rank.}
\label{fig:tpr_white_top_50}
\vspace{1ex}
\end{figure}

We also perform a qualitative evaluation of the end-to-end assignment process using the relevance scoring model. We select six representative (honest) reviewers from our dataset -- Kavita Bala\footnote{\url{https://scholar.google.com/citations?user=Rh16nsIAAAAJ}}, Ryan P. Adam\footnote{\url{https://scholar.google.com/citations?user=grQ_GBgAAAAJ}}, Peter Stone\footnote{\url{https://scholar.google.com/citations?user=qnwjcfAAAAAJ}}, Yejin Choi\footnote{\url{https://scholar.google.com/citations?user=vhP-tlcAAAAJ}}, Emma Brunskill\footnote{\url{https://scholar.google.com/citations?user=HaN8b2YAAAAJ}} and Elad Hazan\footnote{\url{https://scholar.google.com/citations?user=LnhCGNMAAAAJ}} -- representing distinct areas of interest in ML/AI. \cref{tab:case_study} shows the assigned papers for the selected reviewers, which appear to perfectly match the area of expertise for the respective reviewers. Many of the assigned papers have a bid score of 0 despite being very relevant for the reviewer, which shows that the scoring model is able to discover missing bids and improve the overall assignment quality.

\section{Additional Experiment on White-box Attack}
\label{sec:white-box-additional}

In \cref{sec:experiment} we evaluated our defense against white-box attacks that succeeded in securing the target paper assignment. However, in doing so, it is possible that malicious reviewers that did not succeed initially will inadvertent become high-ranked \emph{after} other reviewers are removed from the candidate set. Therefore, it may be necessary to detect \emph{all} attack instances in the candidate set rather than ones that were successfully assigned.

\cref{fig:tpr_white_top_50} shows the detection TPR for all attackers that were initially ranked below $K=50$ but managed to move into the candidate set after the attack. Since this attacker pool includes many that obtained a relatively low rank, detection TPR is much higher than that of \cref{fig:tpr}. For instance, for $M_d=5$, even when the colluding party is significantly larger at $M_a=10$, detection remains viable with a TPR of more than 40\%. This experiment shows that our detection mechanism is unlikely to inadvertently increase the success rate of failed attacks.

\section{Black-box Attack}
\label{sec:black-box}

The white-box attack from \cref{sec:cheating} assumed that the adversary has extensive knowledge about the assignment system and all reviewers' features/bids. In this section, we propose a more realistic \emph{colluding black-box attack}, where the adversary only has access to the features/bids of reviewers in the colluding party. This attack represents a reasonable approximation of what a real world adversary could achieve, and we show that it is potent against the scoring model in \cref{sec:model} absent of any detection mechanism.

\paragraph{Colluding black-box attack.} The failure of the simple black-box attack from \cref{sec:motivation} is due to the malicious reviewer $r$ bidding positively only on a single paper, instead of also on a group of papers that are similar to $p$. We alter the attack strategy by giving the largest bid score to $U=60$ papers $p'$ that are most similar to $p$ (including $p$ itself). In practice, this can be done by comparing the titles and abstracts of $p'$ to the target paper $p$. We simulate this attack in our experiment by select papers $p'$ whose feature vector $X_{r,p'}$ have a high inner product with $X_{r,p}$.

We can extend this strategy to allow for colluding attacks. The malicious reviewer first selects $M_a-1$ reviewers with the most similar background to form the colluding group. In simulation, we measure reviewer similarity by the inner product between their respective reviewer-related features. Mimicking $r$'s paper selection strategy, every reviewer $r'$ in the colluding group now gives the largest bid score to the $U=60$ papers $p'$ with the highest inner product between $X_{r',p'}$ and $X_{r,p}$.

\begin{figure*}[t!]
\centering
\includegraphics[width=0.6\linewidth]{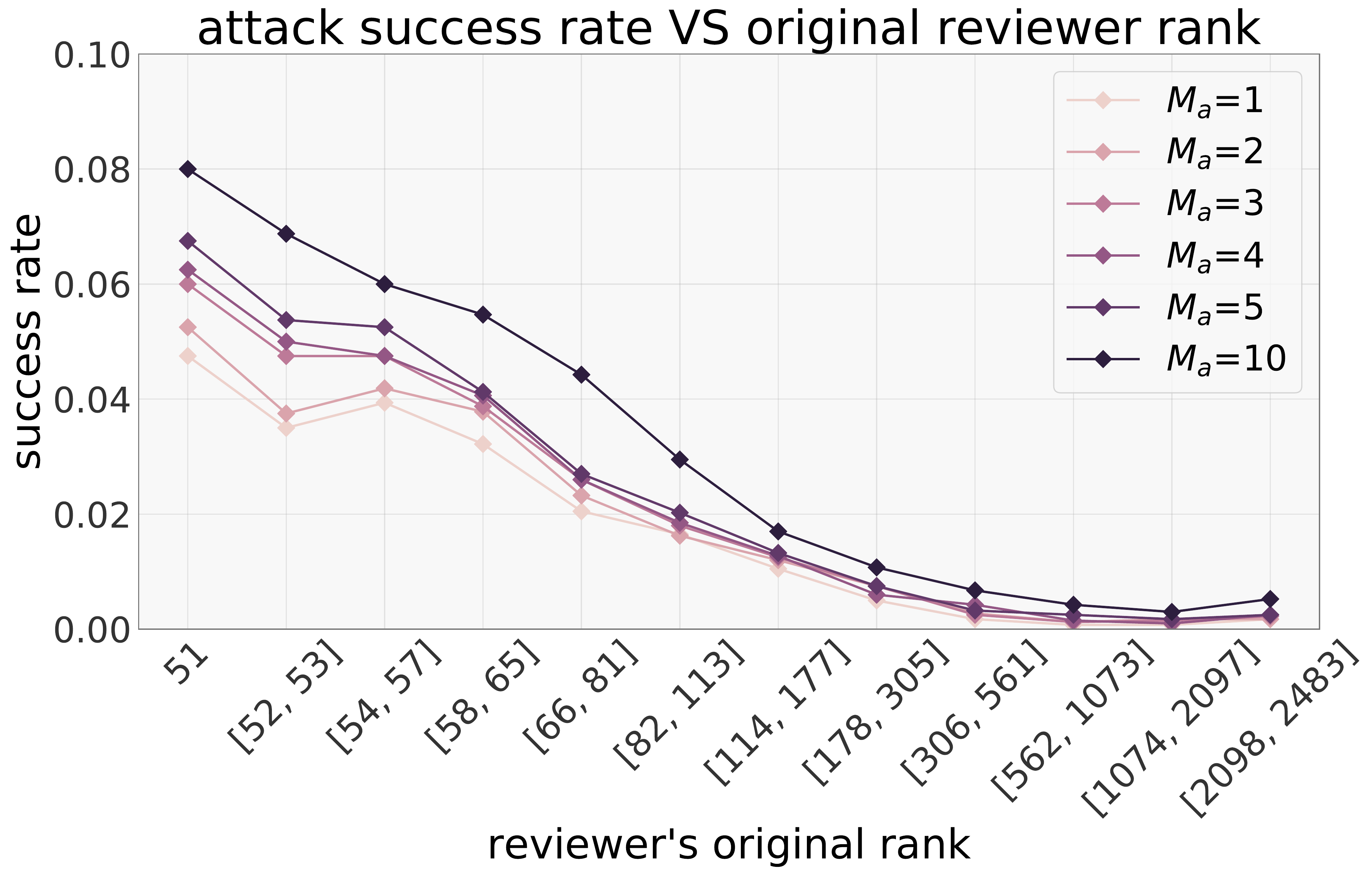}
\caption{Success rate after the \emph{colluding black-box attack} against an undefended linear regression scoring model.}
\label{fig:prob_black}
\end{figure*}

\begin{figure}[t!]
\centering
\includegraphics[width=0.49\textwidth]{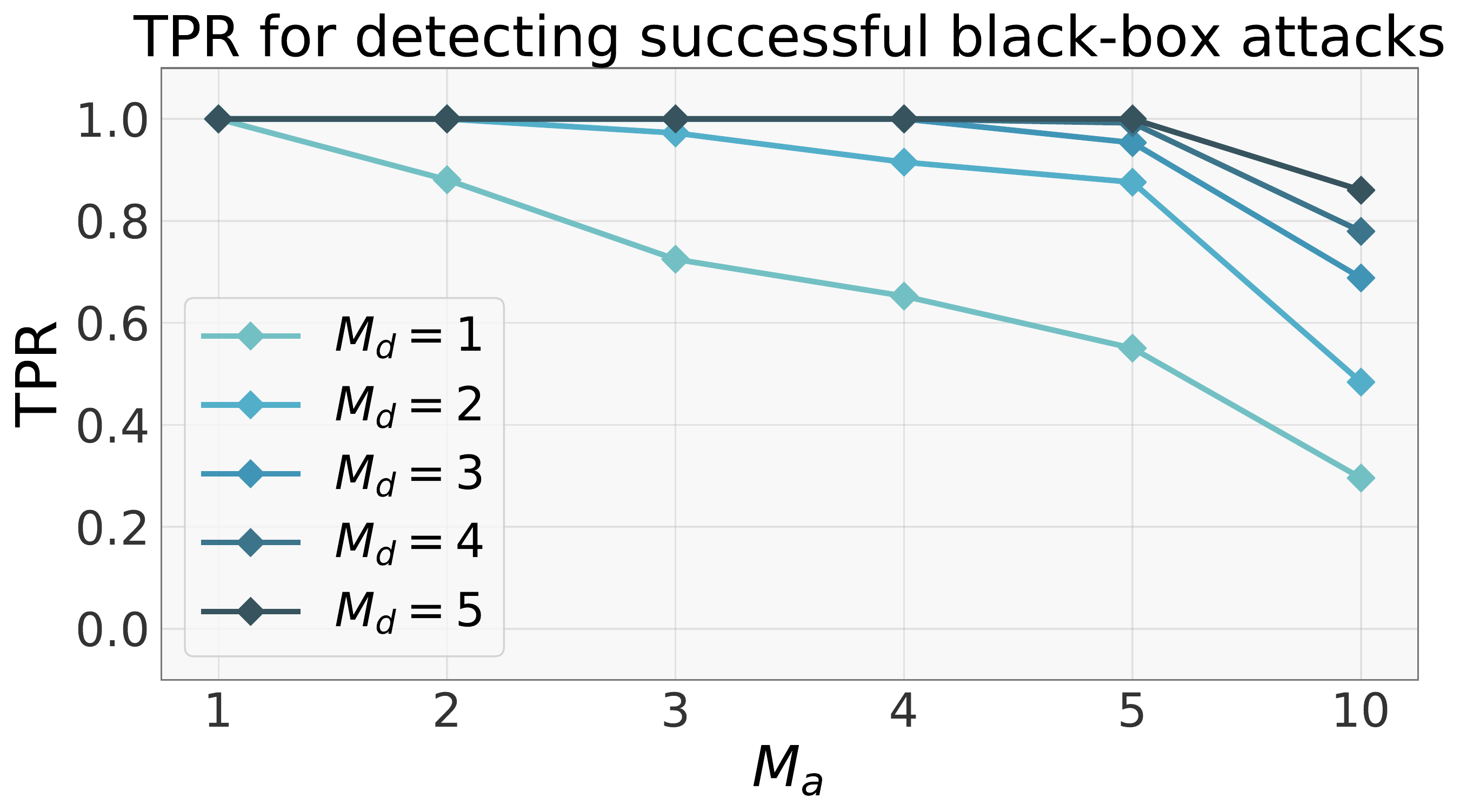}
\includegraphics[width=0.49\linewidth]{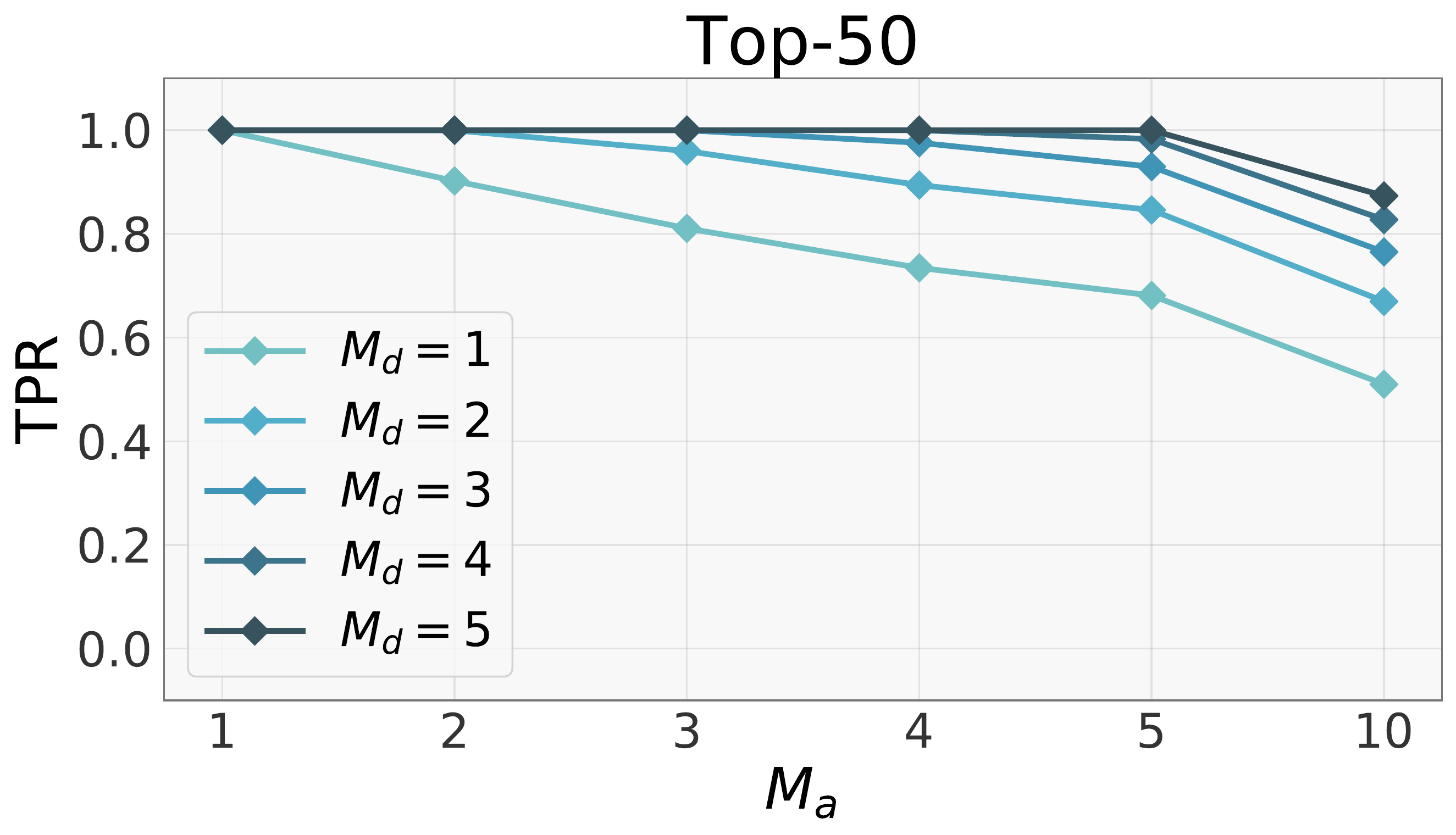}
\vspace{-2ex}
\caption{TPR for detecting \emph{colluding black-box attacks} that succeeded in securing the assignment (left) and achieving a top-50 rank (right).}
\label{fig:tpr_black}
\vspace{1ex}
\end{figure}

\paragraph{Attack performance.} \cref{fig:prob_black} shows the success rate of the colluding black-box attack against the linear regression model. Note that this attack is much more successful than the \emph{simple black-box attack} from \cref{sec:motivation}, which had a success rate of 0\% for all reviewers below rank 16. Here, the success rate before attack is initially 0\%, which increased to close to 5\% after attack even without collusion ($M_a=1$). Increasing the colluding party size strictly improves attack performance, while attackers with lower initial rank are less successful. Compared to the white-box attack from \cref{sec:cheating} (see \cref{fig:prob}), the colluding black-box attack is substantially less potent as expected.

\paragraph{Detection performance.} For completeness, we evaluate the detection algorithm from \cref{sec:detection} against successful colluding black-box attacks. In \cref{fig:tpr_black}, we plot detection TPR as a function of the size of the colluding party ($M_a$) for various choices of the detection parameter $M_d$. For both attacks that succeeded (left) and ones that achieved a top-50 (right) rank, detection TPR is close to 1 when $M_a \leq M_d$, and remains very high for $M_a > M_d$. For instance, at $M_a = 10$ and $M_d = 5$, detection TPR is above 80\% for successful attacks (left plot), which is in sharp contrast with the same setting in \cref{fig:tpr} for the white-box attack, where TPR is reduced to 0\%. The detection performance against this more realistic colluding black-box attack further validates our robust assignment algorithm as a practical countermeasure against bid manipulation.

\end{document}